\documentclass[twocolumn,pre,english,nofootinbib,floatfix]{revtex4-1}

\pdfoutput=1

\usepackage{graphicx}
\usepackage{amsmath}
\usepackage{subfigure}
\usepackage{babel}
\usepackage{array}

\newcommand{\pd}[2]{\frac{\partial#1}{\partial#2}}

\begin{document}

\title{Temperature and Length Scale Dependence of Solvophobic Solvation in a
  Single-site Water-like Liquid}

\author{John R. Dowdle}
\affiliation{The Dow Chemical Company, Freeport TX, 77541, USA}

\author{Sergey V. Buldyrev}
\affiliation{Department of Physics, Yeshiva University, New York, NY 10033 USA}

\author{H. Eugene Stanley}
\affiliation{Center for Polymer Studies and Department of Physics,
Boston University, Boston, MA 02215 USA}

\author{Pablo G. Debenedetti}
\affiliation{Department of Chemical and Biological Engineering, Princeton
  University, Princeton, New Jersey, 08544 USA}

\author{Peter J. Rossky}
\affiliation{Institute for Computational Engineering and Sciences and
  Department of Chemical Engineering, The University of Texas at Austin,
  Austin, Texas 78712, USA}

\begin{abstract}
  The temperature and length scale dependence of solvation properties of
  spherical hard solvophobic solutes is investigated in the Jagla liquid, a
  simple liquid that consists of particles interacting via a spherically
  symmetric potential combining a hard core repulsion and a longer ranged soft
  core interaction, yet exhibits water-like anomalies.  The results are
  compared with equivalent calculations for a model of a typical atomic liquid,
  the Lennard-Jones (LJ) potential, and with predictions for hydrophobic
  solvation in water using the cavity equation of state and the extended simple
  point charge (SPC/E) model. We find that the Jagla liquid captures the
  qualitative thermodynamic behavior of hydrophobic hydration as a function of
  temperature for both small and large length scale solutes.  In particular,
  for both the Jagla liquid and water, we observe temperature-dependent
  enthalpy and entropy of solvation for all solute sizes as well as a negative
  solvation entropy for sufficiently small solutes at low temperature.  This
  feature of water-like solvation is distinct from the strictly positive and
  temperature independent enthalpy and entropy of cavity solvation observed in
  the Lennard-Jones fluid. The results suggest that, compared to a simple
  liquid, it is the presence of a second thermally accessible repulsive energy
  scale, acting to increasingly favor larger separations for decreasing
  temperature, that is the essential characteristic of a liquid that favors
  low-density, open structures and models hydrophobic hydration, and that it is
  the presence of this second energy scale that leads to the similarity in the
  behavior of water and the Jagla liquid. In addition the Jagla liquid dewets
  surfaces of large radii of curvature less readily than the Lennard-Jones
  liquid, reflecting a greater flexibility or elasticity in the Jagla liquid
  structure than that of a typical liquid, a behavior also similar to that of
  water's hydrogen bonding network.  The implications of the temperature and
  length scale dependence of solvation free energies in water-like liquids are
  explored with a simple model for the aggregation of solvophobic solutes.  We
  show how aggregate stability depends upon the size of the aggregate and the
  size of its constituent solutes, and we relate this dependence to
  cold-induced destabilization phenomena such as the cold-induced denaturation
  of proteins.
\end{abstract}

\maketitle

\newpage

\section{Introduction}

Among the many anomalous properties of liquid water is the solvation behavior
of small apolar solutes, which is characterized at ambient conditions by an
unfavorable entropy of transfer from vapor phase to water and an atypical
decrease in solubility with increasing temperature.  This behavior contrasts
with typical solvents, which more readily accommodate apolar compounds as
thermal fluctuations increase.  The enthalpy of transfer for non-polar solutes
to low-temperature water is actually negative and favorable, but the solubility
is dominated by the entropic penalty.  These characteristics change as a
function of temperature and solute size.  At sufficiently high temperatures the
enthalpy is large and unfavorable and is only partially compensated for by
favorable transfer entropies.  Similarly, for sufficiently large solutes, the
poor solubility is dominated by the unfavorable enthalpy associated with the
formation of an interface, which overcomes the favorable entropy gain
\cite{chandler2005interfaces}.

Recent theoretical work in the field of hydrophobic solvation
\cite{lum1999hydrophobicity,huang2001scaling,huang2002hydrophobic,huang2000cavity,rajamani2005hydrophobic,patel2010fluctuations}
has refocused attention on the size-dependence of solvation free energy for
small and large solutes, which is generally accepted to play a potentially
important role in the formation and stabilization of many biological structures
including proteins and cell membranes.  Specifically, it was demonstrated that
the solvation free energies of simple hard sphere solutes in water at ambient
conditions undergo a crossover in size dependence at about 1 nm
\cite{chandler2005interfaces}.  For solutes of size smaller than 1 nm, the
solvation free energy scales with the volume of the solute, while for larger
solutes it scales with the surface area.  This crossover behavior is general to
all liquids far from the critical point and near liquid-vapor coexistence, but
the length scale of the crossover in water is greater than that of simple
liquids, such as a simple Lennard-Jones (LJ) liquid \cite{huang2001scaling}.
This longer crossover distance is attributed to water's propensity to create
available space throughout its hydrogen bonding network.

Traditional explanations of hydrophobic behavior, and water-like anomalies in
general, place emphasis on the orientational interactions of water molecules
(hydrogen bonding) and the accompanying tendency for tetrahedral structure.
However, it has been demonstrated
\cite{yan2005structural,yan2006family,buldyrev_unusual_2009} that water-like
thermodynamic and structural anomalies can also be manifested by a recently
introduced family of spherically symmetric potentials which possess two
characteristic length scales (the Jagla model
\cite{jagla1998phase,jagla1999core}), a hard core and a longer ranged soft core
repulsion.  Further, the Jagla model has also been shown to exhibit water-like
solvation thermodynamics \cite{buldyrev2007water}.  In particular, the
solubility of simple hard sphere solutes in the Jagla liquid is a non-monotonic
function of the temperature, and furthermore, a polymer composed of such hard
spheres exhibits a solvent-induced collapsed state with a stability diagram in
the pressure-temperature plane reminiscent of that of a typical globular
protein in water
\cite{buldyrev2007water,buldyrev_hydrophobic_2010,maiti_potential_2012}.  These
results confirm that orientational interactions are not necessary to produce
these features of water-like solvation behavior
\cite{molinero2009intermediate,moore2010ice,moore2011structural} and suggest
that the presence of two competing length scales is a fundamental physical
feature of hydrophobic hydration.

Questions still remain, however, about the similarities between solvation in
the Jagla liquid and water.  In particular, what are the energetic and entropic
contributions to the solvation free energy in the Jagla liquid and are they
similar to those of water?  Over what length scales do the analogies in
solvation behavior between the two liquids extend?  Is the length scale
crossover behavior in the Jagla liquid similar to that of other simple liquids,
or does it also mimic that of water?  In the present study, we address all of
these questions using extensive Monte Carlo (MC) simulations of the Jagla
liquid. In addition, we compare results for water and the Jagla liquid to
results for the LJ liquid wherever possible.  In doing so we clarify what is
indeed unique to water-like solvation and what is common in typical liquids.


This paper is organized as follows.  In Section~\ref{sec:methods}, we describe
the theoretical and computational methods used to calculate the thermodynamic
quantities of interest.  In Section~\ref{sec:models}, we describe the
interparticle potentials used and the details of the simulation protocols.  The
results of the calculations are presented and discussed in
Sec. \ref{sec:results}, and conclusions and future directions are given in
Sec. \ref{sec:conclusions}.

\section{Theoretical \& Computational Methods}
\label{sec:methods}

All solvation properties of a solute may be obtained once the excess chemical
potential is known. Thus, our calculations focus on the evaluation of the
excess chemical potential of a cavity solute, $\mu_c^x$, which is formally
given by

\begin{equation}
  \label{eq:mu_ex_small}
  \mu_c^x(R) = - k_B T\ln{p_0(R)},
\end{equation}

where $T$ is the temperature, $k_B$ is Boltzmann's constant, and $p_0(R)$ is
the probability of finding a cavity of size $R$ or larger around a randomly
located point in solution. For sufficiently small cavities, $p_0(R)$ may be
evaluated directly via the test particle insertion method
\cite{widom1963some,widom1982potential}.  In dense liquids, however, the
probability of observing density fluctuations extreme enough to accommodate
cavities much larger than the solvent particles is exceedingly small, and test
particle insertion is known to fail in this case
\cite{frenkel2002understanding}.

There are several methods available for the evaluation of chemical potentials
for large cavities (see \emph{e.g.}, \cite{kofke1997quantitative}), but for the
Jagla and LJ fluids in this study we choose to use the revised scaled particle
theory (RSPT) of Ashbaugh and Pratt
\cite{ashbaugh2006colloquium,ashbaugh2007contrasting}.  Here we give only a
brief overview of RSPT which closely follows that given in
Ref. \cite{ashbaugh2009blowing}.  For more detailed descriptions the reader is
referred to Refs. \cite{ashbaugh2006colloquium,ashbaugh2007contrasting}.

RSPT improves upon classical scaled particle theory (SPT)
\cite{reiss1959statistical,stillinger1973structure} by including multi-body
correlations. The essential idea behind both RSPT and SPT is that the excess
chemical potential must be equal to the work required to inflate a cavity
against the solvent from size zero to $R$. This work must oppose the pressure
due to the solvent molecules at the cavity boundary, and thus scaled particle
theories require knowledge of the contact correlation function, $G(R)$, defined
to be the average density of solvent molecules, relative to the bulk, at the
cavity-solvent interface. With $G(R)$ known, the excess chemical potential is
calculated as

\begin{equation}
  \label{eq:mu_ex_spt}
  \mu_c^x(R) = \int_0^R k_B T\rho G(r) 4\pi r^2 dr,
\end{equation}

where $\rho$ is the bulk solvent number density. For $R$ much greater than the
solvent size, the contact correlation function may be expanded in curvature,
$R^{-1}$, with phenomenological coefficients

\begin{equation}
  \label{eq:GofR_expansion}
  G(R) = \frac{\beta P}{\rho} + \frac{2\beta\gamma_{\infty}}{\rho R} 
  - \frac{4\beta \gamma_{\infty}\delta}{\rho R^2} + \dots
\end{equation}

Here, $P$ is the bulk pressure, $\gamma_{\infty}$ is the surface tension of a
flat solvent-cavity interface, and $\delta$ is the first-order curvature
correction to the surface tension \cite{ashbaugh2009blowing}. An expression
for the excess chemical potential of large cavity solutes is then be obtained
by expanding Eq. (\ref{eq:GofR_expansion}) to fourth order and integrating to
get

\begin{align}
  \label{eq:mu_ex_large}
  \mu_c^x(R)|_{\mathrm{large}} = & \frac{4\pi R^3 P}{3} + 4\pi R^2\gamma_{\infty} - 
  16\pi\gamma_{\infty}\delta R \notag \\
  & + 4\pi k_B T\rho\kappa - \frac{4\pi k_B T\rho\lambda}{R},
\end{align}

where $\lambda$ is the fourth-order curvature correction coefficient and
$\kappa$ is an integration constant. Third order coefficients are typically
set to zero so as to avoid logarithmic contributions to $\mu_c^x$
\cite{tully1970further,stillinger1971free}, a convention we follow in
this work. The results for the test particle insertion calculations for small
cavities, $\mu_c^x(R)|_{\mathrm{sim}}$, are interpolated with the large cavity
solute expression in Eq. (\ref{eq:mu_ex_large}) by

\begin{equation}
  \label{eq:mu_ex_rspt}
  \mu_c^x(R) = \mu_c^x(R)|_{\mathrm{sim}} f(R) + \mu_c^x(R)|_{\mathrm{large}} (1-f(R)).
\end{equation}

The function $f(R)$ used here is a cubic function designed to smoothly switch
between small ($R_{\mathrm{sim}}$) and large ($R_{\mathrm{large}}$) cavity sizes,

\begin{widetext}
\begin{equation}
  f(R) =
  \begin{cases}
    1, & R < R_{\mathrm{sim}}, \\
    1 - 3\frac{(R - R_{\mathrm{sim}})^2}{(R_{\mathrm{large}} - R_{\mathrm{sim}})^2} 
    + 2 \frac{(R-R_{\mathrm{sim}})^3}{(R_{\mathrm{large}} - R_{\mathrm{sim}})^3},
    & R_{\mathrm{sim}} \leq R \leq R_{\mathrm{large}}, \\
    0, & R > R_{\mathrm{large}}.
  \end{cases}
\end{equation}
\end{widetext}

\begin{figure}[htbp]
  \centering
  \includegraphics[width=0.9\columnwidth]{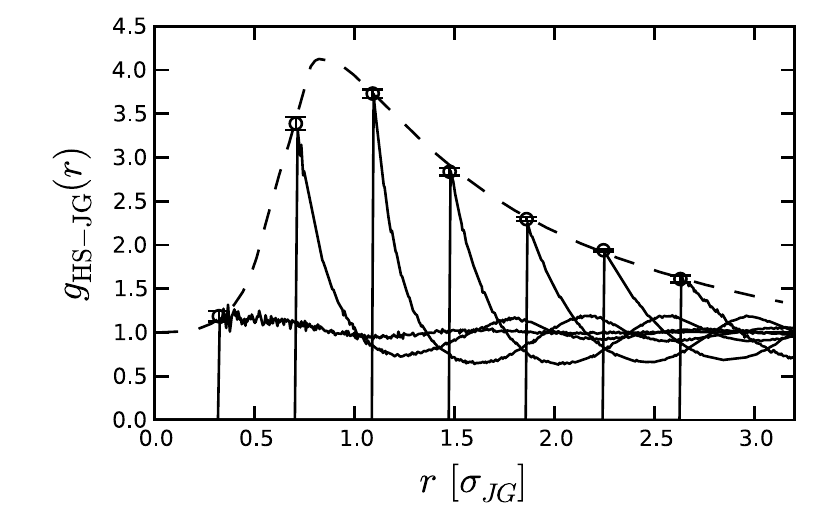}
  \caption{Demonstration of a fit of Eq. (\ref{eq:GofR_rspt}) for the cavity
    contact correlation function to calculated contact values for several
    cavity sizes in the Jagla liquid at $T=0.6$ $[\varepsilon_2/k_B]$. The
    contact correlation function, $G(R)$ (dashed line), is fit to the maxima
    (open circles) in the cavity-solvent pair correlation functions,
    $g_{HS-JG}(r)$ (solid lines). The cavity radii are, in units of
    $\sigma_{JG}$, 0.32, 0.71, 1.09, 1.47, 1.86, 2.24, and 2.63.}
  \label{fig:GofR_with_gofr}
\end{figure}

In order to obtain parameters appearing in the expansion for the contact
correlation function, we use Eq. (\ref{eq:mu_ex_rspt}) and differentiate
Eq. (\ref{eq:mu_ex_spt}) with respect to $R$ to obtain the contact correlation
function as

\begin{align}
    G(R) = & \frac{f(R)}{4\pi \rho R^2} \frac{\partial \beta \mu_c^x(R)|_{\mathrm{sim}}}{\partial R} 
    + \frac{\beta \mu_c^x(R)|_{\mathrm{sim}}}{4\pi \rho R^2} \pd{f(R)}{R} \notag \\
    & + \left( \frac{\beta P}{\rho} + \frac{2\beta\gamma_{\infty}}{\rho R} 
      - \frac{4\beta\gamma_{\infty}\delta}{\rho R^2} + \frac{\lambda}{R^4} \right)
    \left[1 - f(R)\right] \notag \\
    &- \left( \frac{\beta PR}{3\rho} + \frac{\beta\gamma_{\infty}}{\rho} 
      - \frac{4\beta\gamma_{\infty}\delta}{\rho R} + \frac{\kappa}{R^2} 
      - \frac{\lambda}{R^{3}} \right)\frac{\partial f(R)}{\partial R},
    \label{eq:GofR_rspt}
\end{align}

and fit this function to the contact values calculated from the MC simulations,
as demonstrated in Fig. \ref{fig:GofR_with_gofr}. The pressure is set equal to
the simulation pressure, and the parameters $\gamma_{\infty}$, $\delta$,
$\kappa$, and $\lambda$ are fit to the simulation results.

The contact density calculations for the Jagla and LJ liquids demand
significant amounts of computer time to obtain good statistics, and performing
similar calculations for typical multi-site water models that have
electrostatic interactions is not desirable. For our purposes of comparison
here, we may, however, estimate the excess chemical potential of large cavities
in water over a broad range of thermodynamic states by using the recently
developed cavity equation of state (C-EoS) \cite{ben2005global}. The C-EoS is
an analytical equation of state parameterized to fit experimental and
simulation results for water, and it has been shown to accurately reproduce
hydrophobic solvation thermodynamics of simple hydrophobes when combined with a
first-order perturbation theory. The functional form of the C-EoS is given by

\begin{equation}
  \label{eq:ceos_generlal_form}
  \beta\mu_c^x = a + b \beta + c \ln{\beta},
\end{equation}

where $\mu_c^x$ is the cavity chemical potential and the coefficients, $a$,
$b$, and $c$ are assumed to be temperature independent. Thus, the C-EoS assumes
that the enthalpy of cavity formation depends linearly upon temperature and
that the associated heat capacity is temperature independent. The dependence of
$\mu_c^x$ on the cavity size, $R$, is obtained by expanding in powers of $1/R$
and requiring that $\beta\mu_c^x$ approach $\gamma_{lv}a_0$ in the large cavity
limit, where $\gamma_{lv}$ is the experimental liquid-vapor surface tension and
$a_0=4\pi R^2$ is the cavity surface area,

\begin{align}
  \label{eq:ceos}
  \beta\mu_c^x / a_0 = & \sum_{i=0}^3 A_i (1/R)^i + \left[ \sum_{i=0}^3 B_i
    (1/R)^i \right]\beta \notag \\
  & + \left[ \sum_{i=0}^3 C_i (1/R)^i \right]\ln{\beta}.
\end{align}

The remaining coefficients $A_i$, $B_i$, and $C_i$ are obtained from fits to
simulation data.

\section{Simulation Details}
\label{sec:models}

MC simulations of cavity solvation in the Jagla and LJ fluids were performed
along the liquid vapor coexistence curves of each liquid for states ranging
from the triple point to slightly below the critical point.  The Jagla
potential is given by

\begin{equation}
  \label{eq:jagla_potential}
  u_{JG} (r) =
  \begin{cases}
    \text{$\infty$,} & \text{ $r < r_0$},\\
    \text{$m_1r+b_1$,} & \text{ $r_0 < r \leq r_1$},\\
    \text{$m_2r+b_2$,} & \text{ $r_1 < r \leq r_2$},\\
    \text{0,} & \text{ $r > r_2$,}
  \end{cases}
\end{equation}

where

\begin{align}
  \label{eq:jagla_potential_params}
  & m_1 = \frac{-(\varepsilon_2+\varepsilon_1)}{r_1-r_0}, \\
  & b_1 = -\varepsilon_2 - m_1r_1, \\
  & m_2 = \frac{\varepsilon_2}{r_2-r_1}, \\
  & b_2 = -\varepsilon_2 - m_2r_1.
\end{align}

\begin{figure}[htbp]
  \centering
  \includegraphics[width=0.9\columnwidth]{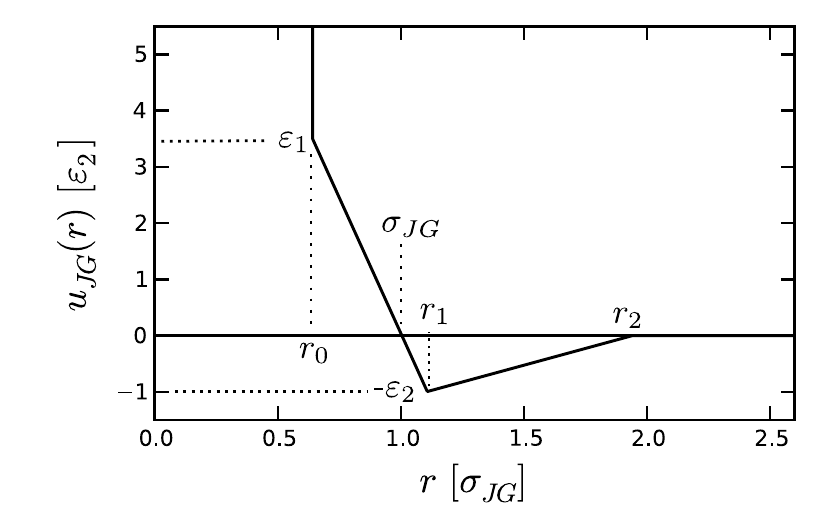}
  \caption{The Jagla two-ramp potential.  The parameters used in the present
    studies are the same as in \cite{buldyrev2007water}, \emph{viz}.:
    $r_1=1.72r_0$, $r_2=3.0r_0$, and $\varepsilon_1=3.5\varepsilon_2$.  The
    relative values of the hard core ($r_0$) and the soft core ($r_1$)
    positions roughly correspond to the same ratio between the positions of the
    first and second solvation shells of liquid water.  The effective size of
    the Jagla particle, $\sigma_{JG}$, is estimated from plots of the radial
    distribution to be the minimum separation at which $u_{JG}(r)=0$ (see
    Fig. \ref{fig:rdfs}).}
  \label{fig:jagla_potential}
\end{figure}

This potential, shown in Fig. \ref{fig:jagla_potential}, demonstrates a wide
range of behavior for varying parameters, including limiting cases of hard
sphere, triangle well, and ramp potentials.  Here we choose $r_1 = 1.72r_0$,
$r_2 = 3.0r_0$, and $\varepsilon_1 = 3.5\varepsilon_2$, as this particular
parameterization manifests a cascade of water-like anomalies
\cite{errington2001relationship,buldyrev2007water,xu2006thermodynamics,yan2006family}.

For the LJ fluid we use the cut-shifted LJ interaction given by

\begin{equation}
  \label{eq:lj_potential}
  u_{LJ}^{cut} (r) =
  \begin{cases}
  u_{LJ}(r) - u_{LJ}(r_c), & r < r_c, \\
  0,   & r \geq r_c,
  \end{cases}
\end{equation}

where $u_{LJ}(r) = 4\varepsilon_{LJ}\left(\sigma_{LJ}^{12}/r^{12} -
  \sigma_{LJ}^6/r^6\right)$ is the full LJ interaction, $\varepsilon_{LJ}$ and
$\sigma_{LJ}$ are the well depth and solvent diameter, respectively, and the
cutoff distance, $r_c$, used is chosen as $2.5\sigma_{LJ}$.


Several different sets of Monte Carlo simulations were performed on the Jagla
liquid.  In the first, saturation properties of the Jagla fluid were estimated
from canonical ensemble MC simulations of a liquid slab in equilibrium with its
vapor for selected temperatures ranging from near the triple point to just
below the critical point.  From these slab simulations we estimate saturated
liquid and vapor densities, the saturation pressure, and the liquid-vapor
surface tension along the binodal line.  The surface tension, $\gamma_{lv}$, is
calculated from the profiles of the pressure tensor using the mechanical
definition \cite{kirkwood1949statistical,walton1983pressure}.  The results for
the saturation properties are shown in Table~\ref{tab:saturation_properties}.


In the second set of simulations, isothermal-isobaric MC simulations of the
Jagla fluid were performed for both the liquid and vapor phases at each of the
saturation states listed in Table~S1 in the supplementary material.  Test
particle insertion calculations were performed on the resulting liquid phase
trajectories for cavities up to 2$\sigma_{JG}$ in diameter to obtain
$\mu_c^x(R)|_{\mathrm{sim}}$.  Similarly, insertion probabilities and excess
chemical potentials for cavities up to 6$\sigma_{JG}$ in diameter were obtained
from test particle insertion analysis of the vapor phase trajectories.
Knowledge of the vapor phase chemical potentials allows evaluation of the
surface tension at the vapor wall interface \cite{ashbaugh2009blowing}.


Finally, isothermal-isobaric MC simulations of a single cavity in the Jagla
liquid were performed for various cavity sizes at each of the saturation states
listed in Table~S1 in the supplementary material. Cavity diameters up to
6$\sigma_{JG}$ were considered, and the contact correlation function was
evaluated for each cavity at each state point.  The contact correlation
function is determined by extrapolating the cavity-solvent pair correlation
function to contact. 

The parameters in Eq. (\ref{eq:GofR_rspt}) may be fit to the MC results for
$G(R)$, and the cavity excess chemical potential may then be computed from
Eq. (\ref{eq:mu_ex_rspt}).  The details of the MC simulations used to calculate
the insertion probabilities and contact correlation functions in the Jagla
fluid are provided in Tables~S1 and S2 in the supplementary material.  All data
for the LJ liquid are those obtained in the studies reported in
Ref. \cite{ashbaugh2009blowing}.  The saturation states for the LJ liquid are
also listed in Table~S4 in the supplementary material for the present study.

Molecular dynamics simulations of the SPC/E water model
\cite{berendsen1987missing} were performed along the liquid vapor coexistence
curve for each of the states listed in Table~S5 in the supplementary material.
A system consisting of 512 SPC/E water molecules was simulated in a cubic box
with periodic boundary conditions in the canonical ensemble for 20 ns using the
GROMACS molecular dynamics engine \cite{berendsen1995gromacs,van2005gromacs}.
The time step was chosen as 2 fs, and bonds were constrained with the SETTLE
algorithm \cite{miyamoto1992settle}.  The velocity rescaling thermostat was
used to control temperature with a time constant of 0.1 ps
\cite{bussi2007canonical}.  Particle mesh Ewald summation was used to treat
long range electrostatic interactions \cite{essmann1995smooth} with a real
space cutoff of 1.2 nm and a mesh spacing of 0.18 nm.  The Ewald tolerance was
set to $10^{-5}$, and fourth order interpolation was used.

\section{Results and Discussion}
\label{sec:results}

\subsection{The Definition of Solvent Size from Pair Distribution Functions}

\begin{figure}[htbp]
  \centering
  \includegraphics[width=0.9\columnwidth]{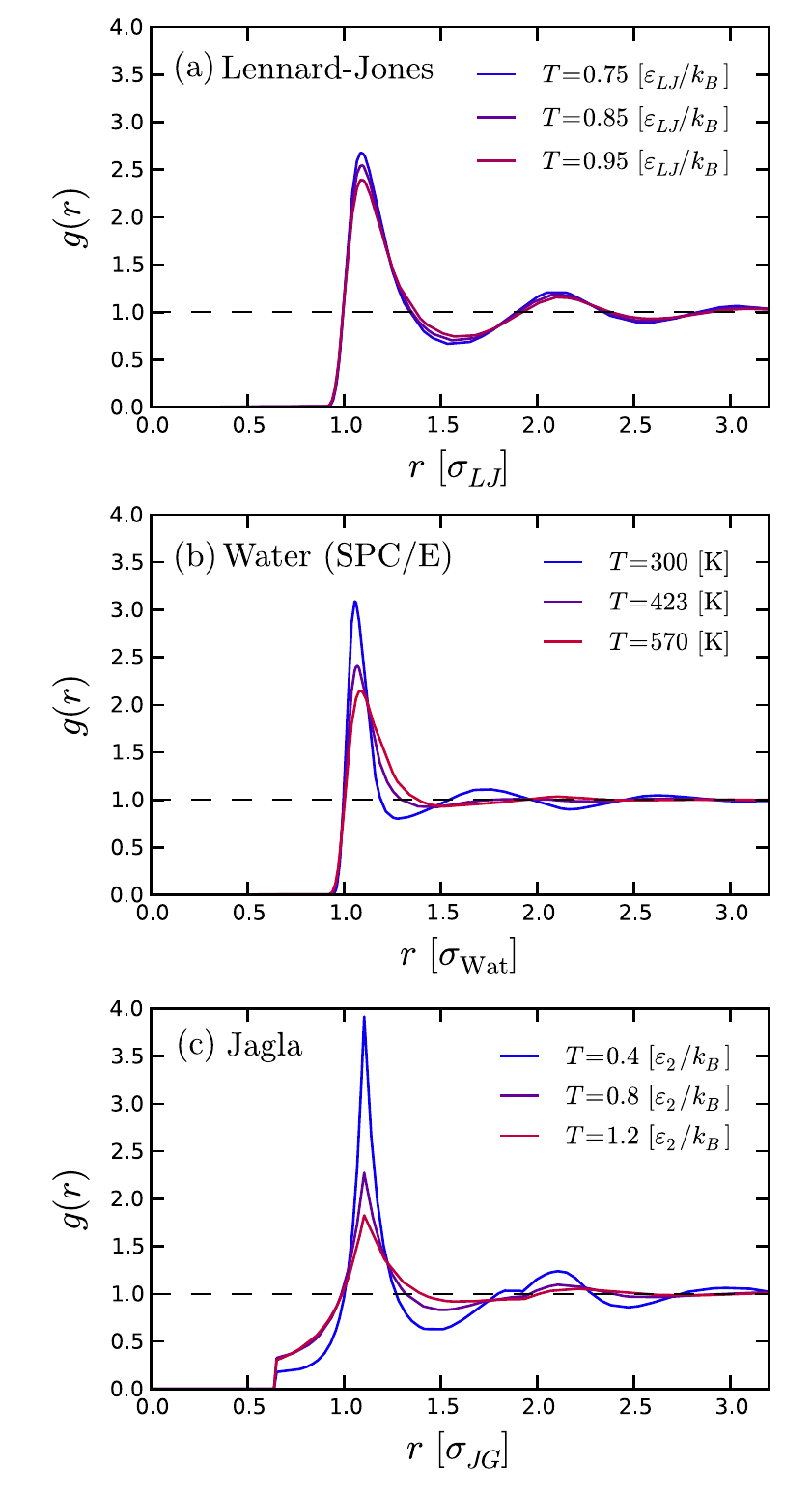}
  \caption{Solvent-solvent pair distribution functions for states along the
    saturation curves of (a) the LJ liquid, (b) SPC/E water, and (c) the Jagla
    liquid. It is evident from the figure that the minimum separation at which
    $g(r)$ has the value unity can be used as an estimate for the solvent
    size. For the SPC/E model this corresponds to $\sigma_{Wat} = 0.26$ nm, for
    the LJ liquid it is $\sigma_{LJ}$, and for the Jagla liquid it is
    $\sigma_{JG}=1.56r_0$ (the minimum separation at which
    $u_{JG}(r)=0$). These sizes are taken to be independent of temperature for
    the states considered here, as justified by the data shown.}
  \label{fig:rdfs}
\end{figure}

A comparison of the solvent-solvent pair correlation function, $g(r)$, for the
three liquids is shown in Fig. \ref{fig:rdfs}. The maximum in $g(r)$ for the LJ
liquid occurs at a pair separation slightly larger than $\sigma_{LJ}$, and at a
separation of $\sigma_{LJ}$ the pair distribution function assumes a value of
very nearly one for all states on the saturation curve.  The nearest separation
at which $g(r)$ is unity is a commonly used estimate for the size of a particle
since the surrounding fluid is depleted from all shorter distances.  We adopt
this estimate here and use $\sigma_{LJ}$ as the size of the LJ particle.

In the case of SPC/E water, the pair distribution function peaks at about 0.28
nm at ambient temperature and slightly larger distances at higher temperatures.
These distances are smaller than the LJ diameter for oxygen due to H-bonding.
The nearest separation at which $g(r)$ is unity is nearly constant at about
0.26 nm, which, to be consistent, is our choice for the size of the SPC/E
molecule, $\sigma_{Wat}$.

The maximum peak in the Jagla liquid $g(r)$ occurs at a distance significantly
larger than the hard core diameter, $r_0$.  This reflects the preference of
Jagla particles to maintain separation at the minimum in $u_{JG}(r)$, $r_1$,
unless stressed by temperature or pressure.  This preference is diminished as
temperature increases.  However, the minimum separation at which the Jagla
$g(r)$ is unity is found to be insensitive to temperature [see
  Fig. \ref{fig:rdfs} (c)] and closely corresponds to the minimum separation at
which the pair potential is zero.  This distance, $\sigma_{JG}$, is a
consistent estimate for the size of the Jagla particle; $\sigma_{JG}=1.56r_0$
for the potential parameterization considered here.

\subsection{Surface Tension and Vapor-Liquid Equilibria in the Jagla fluid}

In the first set of MC simulations, saturation properties of the Jagla fluid
were estimated from canonical ensemble MC simulations of a liquid slab in
equilibrium with its vapor for selected temperatures ranging from near the
triple point to below the critical point. From these simulations we estimate
liquid and vapor densities, the saturation pressure, and the liquid-vapor
surface tension along the binodal line.

The results for the liquid-vapor slab simulations of the Jagla fluid are
summarized in Table~\ref{tab:saturation_properties}. The saturated liquid
densities and the equilibrium vapor densities are in close agreement with those
reported by Lomba \emph{et al.} \cite{lomba2007phase}. We expect that our
estimates of the coexistence properties of the Jagla fluid may be improved upon
by taking finite size effects into account, as it is known, \emph{e.g.}, that
large wavelength fluctuations may be suppressed by the system size
\cite{binder1982scaling}. Nevertheless, the solvation behavior we seek to
characterize occurs for states at or near coexistence \cite{huang2001scaling},
and we therefore expect the present estimates from the slab simulations to
suffice for this study.

\begin{table}[!ht]
  \centering
  \begin{ruledtabular}
  \begin{tabular}{cccccc}
    $N$ & $T$ [$\varepsilon_2/k_B$] & $\rho_l$ [$r_0^{-3}$] & $\rho_v$
    [$r_0^{-3}$] & $P_{sat}$ [$\varepsilon_2/r_0^{3}$] & $\gamma_{lv}$
    [$\varepsilon_2/r_0^2$] \\ 
    \hline
    1374 & 0.4   & 0.256(2)   & 5(3)$\times 10^{-5}$     & 3(2)$\times 10^{-5}$   & 0.491(8) \\ 
    1374 & 0.6   & 0.255(2)   & 2.3(7)$\times 10^{-4}$   & 1.4(4)$\times 10^{-4}$   & 0.407(7) \\ 
    1386 & 0.8   & 0.244(2)   & 0.0018(2)   & 0.0014(2)   & 0.314(8) \\ 
    1444 & 1.0   & 0.226(3)   & 0.0067(6)   & 0.0056(6)   & 0.213(5) \\ 
    1600 & 1.2   & 0.203(2)   & 0.0174(9)   & 0.015(1)   & 0.115(7) \\ 
  \end{tabular}
  \end{ruledtabular}
  \caption{Canonical ensemble MC simulations of a liquid slab in equilibrium with
    its vapor were performed to obtain estimates of
    saturation properties. $N$ Jagla particles were simulated at five different
    temperatures for $1.6\times 10^6$ MC cycles, where one cycle corresponds to
    $N$ MC moves. The liquid and vapor densities were estimated from ensemble
    averages of the densities in the centers of the liquid and vapor regions,
    respectively. Similarly, the saturation pressure was obtained by
    evaluating the pressure tensor in the center of the vapor region. The 
    liquid-vapor surface tension is calculated using the virial relation 
    \cite{kirkwood1949statistical,walton1983pressure}. Numbers in 
    parentheses are estimates of the statistical error in the last digit of 
    the reported value.}
  \label{tab:saturation_properties}
\end{table}

\subsection{Cavity Solvation Thermodynamics}

The parameters in Eq. (\ref{eq:GofR_rspt}) were fit to the MC results for
$G(R)$ in the Jagla liquid the using a least-squares regression. The choice of
$R_{\mathrm{sim}}$ and $R_{\mathrm{large}}$ used in the fit varied with the
thermodyanmic state. Values of $R_{\mathrm{sim}}$ ranged from 0.5 to
0.6$\sigma_{JG}$ and values of $R_{\mathrm{large}}$ ranged from 0.75 to
0.95$\sigma_{JG}$.  In all cases, $G(R)$ was well represented between
$R_{\mathrm{sim}}$ and $R_{\mathrm{large}}$ by differentiation of
$\mu_c^x(R)|_{\mathrm{sim}}$.  The results of the fit are presented in
Table~\ref{tab:GofR_params}. The surface tension of the flat interface,
$\gamma_\infty$, is higher than the liquid-vapor surface tension measured in
the slab simulations at all temperatures. It should be emphasized that
$\gamma_\infty$ does not strictly correspond to the liquid-vapor surface
tension, but rather to the total interfacial free energy between the solvent
and the cavity which consists of contributions from two interfaces---a liquid
vapor interface between the solvent and vapor film surrounding the cavity and
the vapor-wall interface between the vapor film and the cavity surface.  If the
two interfaces are well separated and not interacting with one another, then
$\gamma_\infty$ is equal to the sum of the liquid-vapor and vapor-wall surface
tensions.  Our simulations are sufficiently far from the critical point that
the vapor-wall surface tensions are negligible for all states considered.
Furthermore, the fitted values of $\gamma_\infty$ were insensitive to varying
the maximum cavity diameter used in the fits between 4$\sigma_{JG}$ and
6$\sigma_{JG}$, suggesting the finite cavity sizes considered here are not to
blame. Therefore the difference between $\gamma_\infty$ and $\gamma_{lv}$ is
likely due to other factors such as the finite-size limitations of our
estimates of $\gamma_{lv}$ or the physical impact of quenched fluctuations at
the solvent-wall interface \cite{ashbaugh2009blowing}. The first order
curvature correction, $\delta$, is negative and decreases with increasing
temperature, also consistent with the results for the LJ liquid.  It should
also be mentioned that here $\delta$ need not correspond to the Tolman length
\cite{tolman1949effect}, but rather is treated as a fitting parameter.  The
parameters $\kappa$ and $\lambda$ are negative for all states and diminish in
magnitude as the critical point is approached.

\begin{table}[!ht]
  \centering
  \begin{ruledtabular}
  \begin{tabular}{ccccc}
    $T$ [$\varepsilon_2 / k_B$] & $\gamma_{\infty}$ [$\varepsilon_2 / r_0^2$] &
    $\delta$ [$r_0$] & $\kappa$ [$r_0^3$] & $\lambda$ [$r_0^4$]  \\
    \hline
    0.4  &   0.55(1)   &     -0.01(2)   &    -8.1(9)   &    -13.2(5) \\
    0.5  &   0.51(1)   &     -0.09(3)   &    -6.6(7)   &    -11.4(5) \\
    0.6  &   0.47(1)   &     -0.18(3)   &    -5.5(4)   &    -10.2(3) \\
    0.7  &   0.43(1)   &     -0.27(4)   &    -4.8(4)   &    -9.3(2) \\
    0.8  &   0.38(1)   &     -0.35(5)   &    -4.0(4)   &    -8.1(4) \\
    0.9  &   0.33(2)   &     -0.45(5)   &    -3.4(5)   &    -7.2(2) \\
    1.0  &   0.28(1)   &     -0.59(6)   &    -2.9(6)   &    -6.5(1) \\
    1.1  &   0.22(2)   &     -0.77(5)   &    -2.5(4)   &    -5.8(1) \\
    1.2  &   0.17(2)   &     -0.93(5)   &    -1.9(2)   &    -4.8(1) \\
  \end{tabular}
  \end{ruledtabular}
  \caption{Parameters from the least-squares fit of Eq. (\ref{eq:GofR_rspt}) to
the contact densities obtained from the simulations in Table~S2 in the
supplementary material. The simulation data was split into several blocks, and
the numbers in parentheses represent an error in the last digit in the fitted
parameter corresponding to one standard deviation of the block averages.}
  \label{tab:GofR_params}
\end{table}

\begin{figure}[htbp]
  \centering
  \includegraphics[width=0.9\columnwidth]{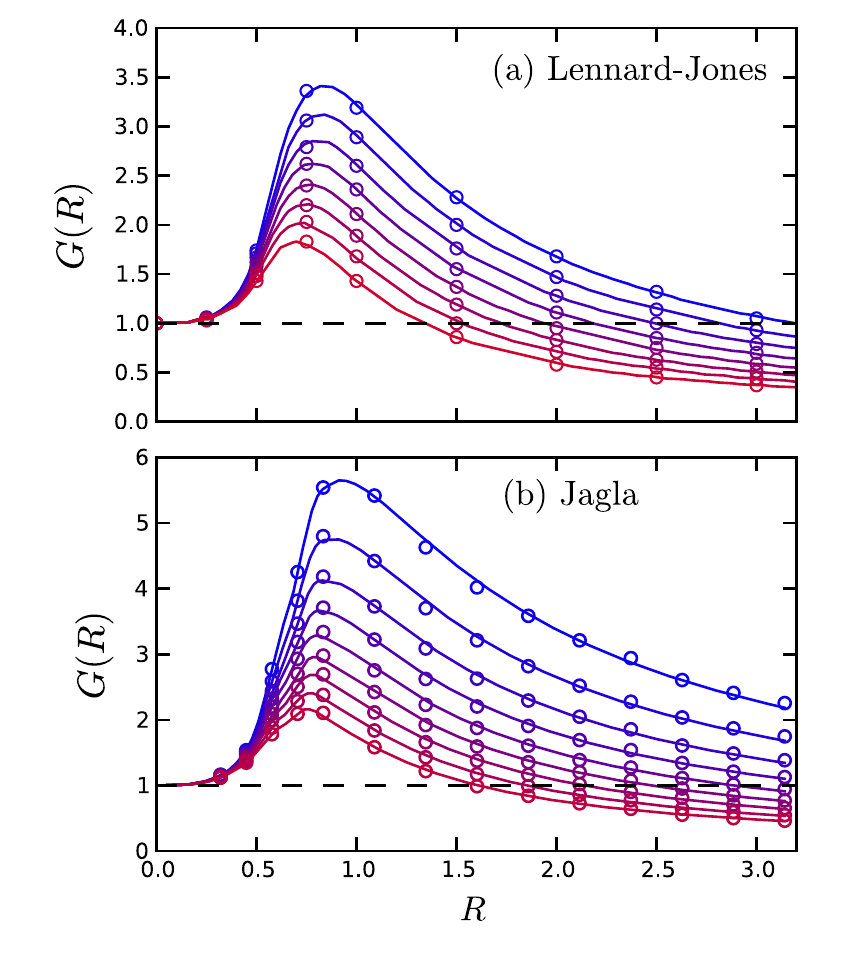}
  \caption{Cavity contact correlation functions as a function of cavity size
    (measured in units of solvent diameters) for states along the saturation
    curves of the (a) LJ and (b) Jagla liquids ranging from near the triple
    point (blue) to just below the critical point (red). The temperatures for
    the LJ liquid range from $k_BT/\varepsilon_{LJ}$ = 0.65 (blue) to 1.00
    (red) in increments of 0.05, while those for the Jagla liquid range from
    $k_BT/\varepsilon_2$ = 0.4 (blue) to 1.2 (red) in increments of 0.1. Points
    are obtained from MC simulation data and lines are fits of
    Eq. (\ref{eq:GofR_rspt}) to the simulation data. Statistical errors are
    smaller than symbol size. All LJ data are obtained from
    Ref. \cite{ashbaugh2009blowing}.}
  \label{fig:GofR_combined}
\end{figure}

The results of the MC calculations for the cavity contact correlation functions
are shown in Fig. \ref{fig:GofR_combined} along with the fits to $G(R)$. In
both fluids, as the solute size grows from zero, the solvent packs increasingly
tightly until the contact density peaks at a value of $R$ on the order of the
solvent size. At this point, the solvent begins to pull away from the solute,
and for sufficiently large solutes, $G(R)$ will be less than one. The contact
correlation function is a decreasing function of temperature for all solute
sizes studied here, but for sufficiently large solute sizes $G(R)$ will
increase with temperature since $\lim_{R\to\infty}G(R)=\beta P / \rho$, which
increases with temperature along the saturation curve.

The cavity sizes where $G(R)$ decreases below one, \emph{i.e.} where the cavity
is ``dewet'', are larger relative to the solvent size in the Jagla liquid,
meaning that the Jagla liquid resists dewetting of hard surfaces more than the
LJ liquid. Lastly, for a fixed cavity size in the LJ liquid the spacing in
$G(R)$ values between temperatures appears roughly constant, suggesting a
linear dependence upon temperature. This is not the case in the Jagla liquid,
however, as the temperature dependence clearly decreases with increasing
temperature.

\begin{figure}[htbp]
  \centering
  \includegraphics[width=0.9\columnwidth]{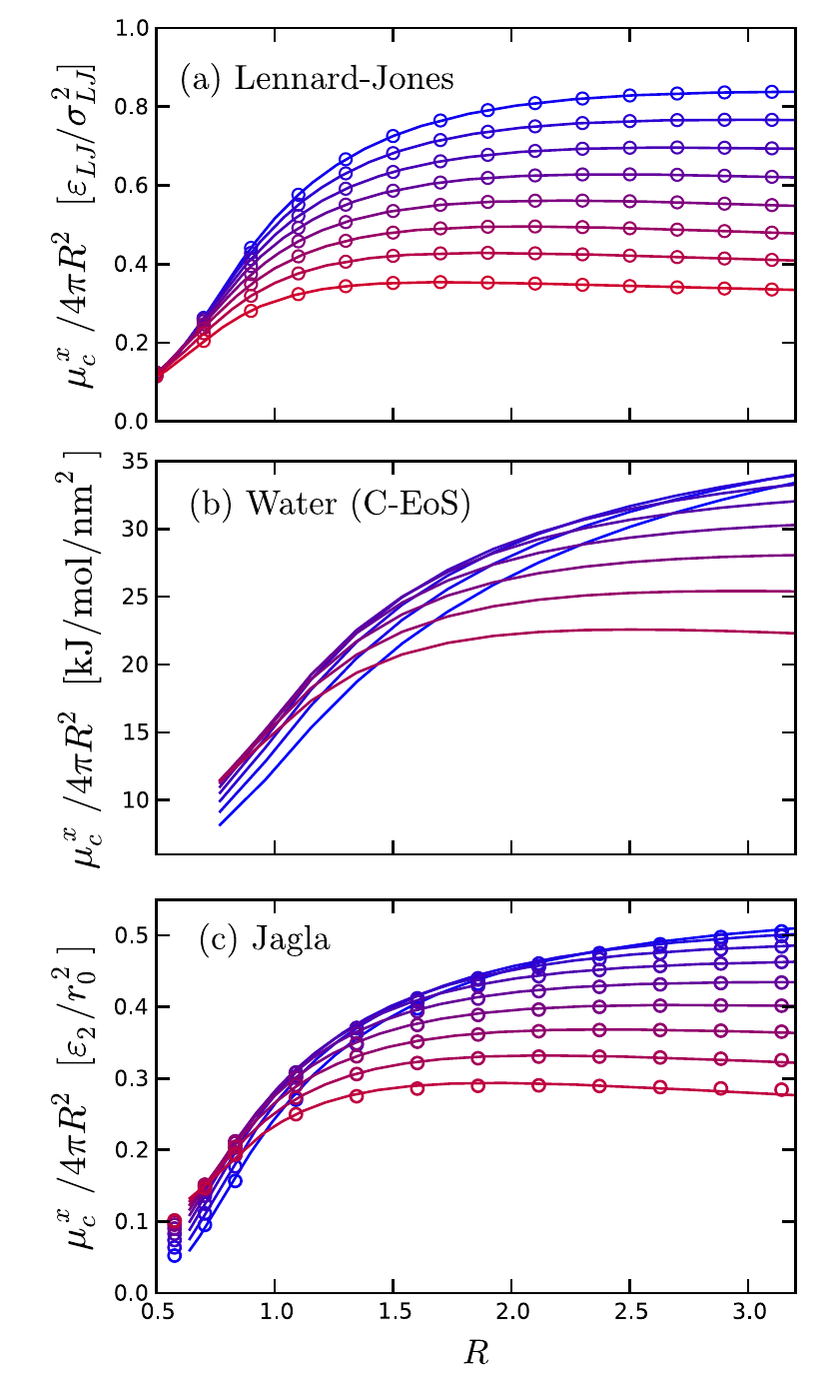}
  \caption{Excess chemical potential per surface area versus cavity size
    (measured in units of solvent diameters) for states along the saturation
    curves of (a) the LJ liquid, (b) water, and (c) the Jagla liquid.  The
    thermodynamic states for the LJ and Jagla liquids are the same as those
    presented in Fig. \ref{fig:GofR_combined}.  Points in the Jagla and LJ
    plots are obtained from simulation data and scaled particle theory.  Lines
    in the LJ plot are fits using Eq. (\ref{eq:lj_mu_ex}), while lines in the
    Jagla plot are fits of the simulation data to the C-EoS
    [Eq. (\ref{eq:ceos})].  Lines in (b) are predictions from the water C-EoS
    \cite{ben2005global}. The temperatures used for the water C-EoS plot are
    $T$ $[K]$ = 273, 304, 335, 366, 398, 429, 460, 491, and 522.}
  \label{fig:mu_ex_per_a}
\end{figure}

With the fitted parameters for $G(R)$, the excess chemical potentials for the
Jagla and LJ liquids may be obtained from Eq. (\ref{eq:mu_ex_rspt}). In the
case of water we use Eq. (\ref{eq:ceos}). The results of the chemical potential
calculations are shown in Fig. \ref{fig:mu_ex_per_a}. The excess chemical
potential is a positive, monotonically increasing function of cavity size at
all temperatures in all three liquids.

In the LJ liquid, the chemical potential is a decreasing function of
temperature for all cavity sizes greater than $\sigma_{LJ}/2$. Furthermore, the
spacing between temperatures for any fixed cavity size appears roughly constant
in the LJ liquid, which, as pointed out by Ashbaugh \cite{ashbaugh2009blowing},
suggests that along the saturation curve the excess chemical potential may be
modeled as

\begin{equation} 
  \mu_c^x(R) = h_c^x(R)|_\sigma - Ts_c^x(R)|_\sigma, 
  \label{eq:lj_mu_ex} 
\end{equation}

where $h_c^x(R)|_\sigma$ and $s_c^x(R)|_\sigma$ are the temperature independent
enthalpy and entropy of solvation. The enthalpy is positive and increases with
cavity size, indicating the loss of favorable solvent-solvent interactions near
the cavity solute. Except for cavities smaller than $\sigma_{LJ}/2$, the
entropy is also a positive, increasing function of cavity size, indicating that
solvent molecules near the cavity experience a net gain in configurational
space. The excellent fit of Eq. (\ref{eq:lj_mu_ex}) to the simulation data
[Fig. \ref{fig:mu_ex_per_a}(a)], indicates that that the enthalpy of solvation
is approximately temperature-independent, and therefore the heat capacity of
cavity solvation in the LJ liquid is approximately zero.  In the Jagla liquid,
in contrast, the chemical potential is an increasing function of temperature
for small, solvent-sized cavities and a decreasing function of temperature for
large cavities.  The temperature derivative of the excess chemical potential
for a fixed cavity size is not constant [Fig. \ref{fig:mu_ex_per_a}(b)], but is
evidently nonlinear. The qualitative behavior of the chemical potential of
cavity solvation in the Jagla liquid is remarkably similar to that predicted
for liquid water by the C-EoS. This suggests that the Jagla liquid data may be
fit to the C-EoS as well. Using the surface tension data
(Table~\ref{tab:saturation_properties}) and a least-squares fit of the excess
chemical potentials calculated from the $G(R)$ data, we obtained a set of C-EoS
parameters for the Jagla liquid (see Table~S5).  The fit is, in fact, excellent
for all cavity sizes and temperatures considered, with slight deviations
occurring only for the largest cavities at the highest temperature. The C-EoS
fit to the simulation data permits exploration of the thermodynamic
contributions to $\mu_c^x$ in the Jagla liquid using analytical derivatives of
Eq. (\ref{eq:ceos}).

\begin{figure*}[htbp]
  \centering
  \includegraphics[width=0.8\textwidth]{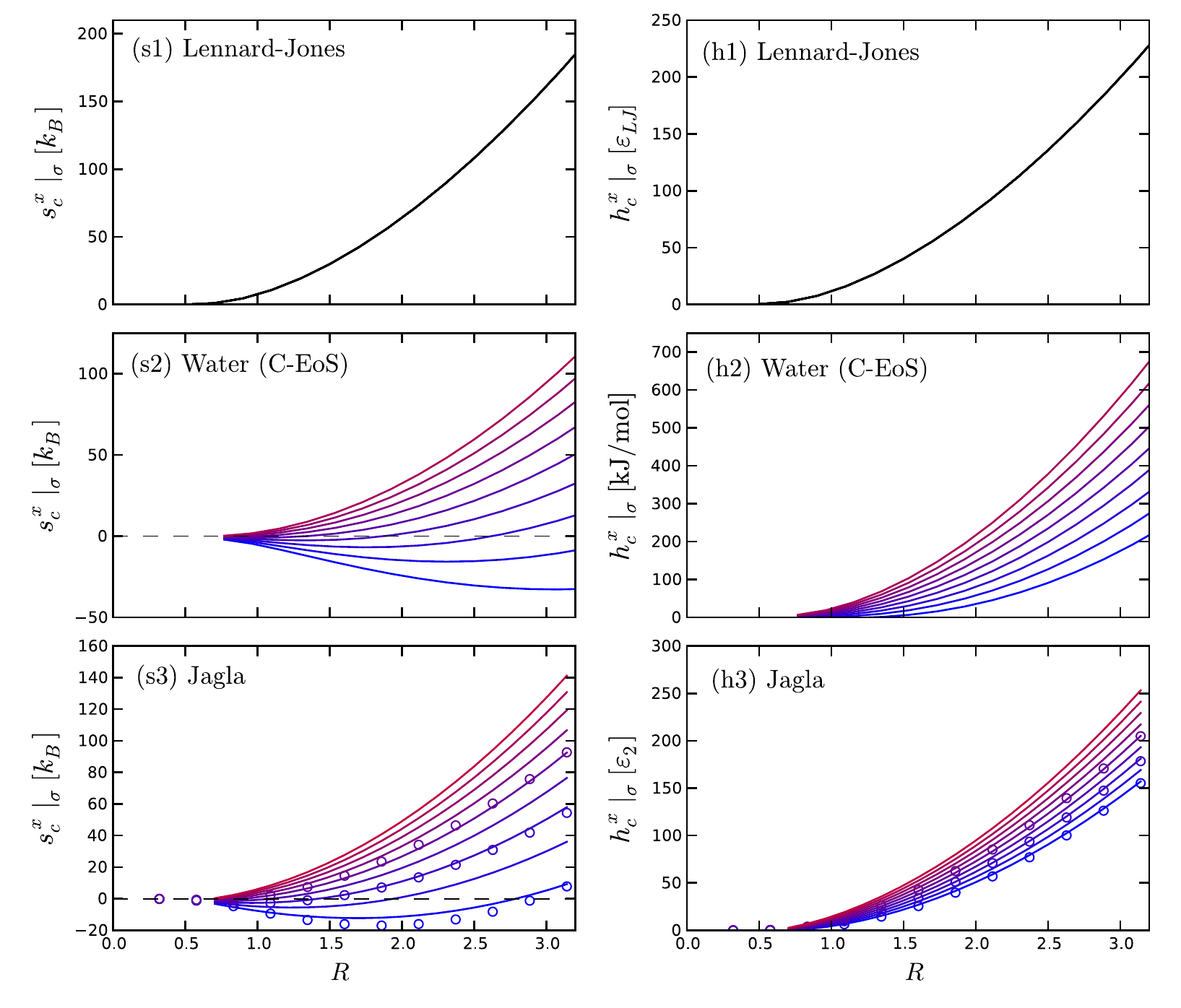}
  \caption{(s1-s3) Entropy and (h1-h3) enthalpy of cavity solvation for the LJ
    liquid, water, and the Jagla liquid as a function of cavity size (measured
    in units of solvent diameters). The temperatures for the Jagla liquid are
    the same as those listed in Fig. \ref{fig:GofR_combined}, while the
    temperatures for the water C-EoS are the same as those listed in
    Fig. \ref{fig:mu_ex_per_a}.  For water and the Jagla liquid, entropies are
    calculated from temperature derivatives of the cavity equation of state
    (lines), $s_c^x|_\sigma=-(\partial{\mu_c^x}/\partial{T})_\sigma$, while for
    the LJ liquid, the entropy is given by the assumed temperature-independent
    form of $\mu_c^x$ in Eq. (\ref{eq:lj_mu_ex}). The enthalpy is calculated
    from $h_c^x|_\sigma=\mu_c^x+Ts_c^x|_\sigma$. Points in (s3) and (h3) are
    numerical derivatives of cubic spline fits to the excess chemical
    potentials in Fig. \ref{fig:mu_ex_per_a}.}
  \label{fig:s_and_h_vs_R}
\end{figure*}

The enthalpic and entropic contributions to the excess chemical potential for
the Jagla liquid and water may be obtained from analytical temperature
derivatives of the C-EoS \footnote{The temperature derivatives are taken along
  the saturation curve, $\sigma$, and they may be related to their constant
  pressure counterparts through the state variable relation
  \cite{reichl2009modern}
\begin{equation*}
  \label{eq:state_variable_relation}
  \left(\pd{\mu_c^x}{T}\right)_\sigma = \left(\pd{\mu_c^x}{T}\right)_P 
  + \left(\pd{\mu_c^x}{P}\right)_T\left(\pd{P}{T}\right)_\sigma.
\end{equation*}
Noting that $(\partial{\mu_c^x}/\partial{P})_T=v_c^x$, where $v_c^x$ is the excess partial molar volume, we may write
\begin{equation*}
  \label{eq:s_ex_ortho}
  s_c^x|_\sigma = -\left(\pd{\mu_c^x}{T}\right)_\sigma = s_c^x|_P - v_c^x\left(\pd{P}{T}\right)_\sigma.
\end{equation*}
Similarly,
\begin{equation*}
  \label{eq:h_ex_ortho}
  h_c^x|_\sigma = \left(\pd{\beta\mu_c^x}{\beta}\right)_\sigma 
  = h_c^x|_P - Tv_c^x\left(\pd{P}{T}\right)_\sigma.
\end{equation*}
The fundamental differences between the liquids considered here are seen in
both the saturation and constant pressure quantities.}. The enthalpy and
entropy of cavity solvation are compared in Fig. \ref{fig:s_and_h_vs_R}.  The
most obvious distinction between the three liquids is that the LJ liquid has
temperature independent enthalpic and entropic contributions to the solvation
free energy, while the contributions for the Jagla liquid and water both show a
strong temperature dependence.  For all three fluids, the enthalpy is a
positive, monotonically increasing function of the cavity radius. The
unfavorable enthalpy results from the disruption of the liquid structure in the
vicinity of the solute and the concomitant formation of an interface which on
average has fewer favorable solvent-solvent interactions than an equivalent
volume in the bulk.

For any fixed cavity size in the size ranges considered in this study, the
enthalpy is an increasing function of temperature in the Jagla liquid and in
water. A possible interpretation for this result in water is given by the
Muller model \cite{muller1990search,graziano2005intactness}, which uses a
simple two-state hydrogen bond (H-bond) model parameterized by empirical
solvation data to argue that the fraction of broken H-bonds in the solvation
shell of apolar solutes is always at least somewhat greater than that in the
bulk, and furthermore, that this disparity increases with temperature. Thus,
for a fixed cavity size an increase in temperature decreases the number of
H-bonds in the solvation shell relative to the bulk, which leads to a greater
enthalpy.

The entropy of cavity formation in both the Jagla liquid and water increases
with increasing temperature for any fixed cavity size. It is possible that this
behavior in water may be also be connected to the breaking of solvation shell
H-bonds. If an increase in temperature causes a decrease in the number of
solvation shell H-bonded pairs relative to bulk, then overall the gain in
configurational freedom will be larger at the higher temperature.  However,
this does not yet explain the Jagla model behavior.

It is remarkable that the Jagla liquid, which contains no orientational
dependence in its interaction potential and therefore no H-bonding, reproduces
the qualitative behavior of hydrophobic hydration thermodynamics.  The
underlying physical origins for this behavior in the Jagla liquid may be
analogous to those of water, however. It has been shown in computer simulations
of SPC/E water that the energetics of H-bonding are strongly correlated with
local crowding effects. In particular, H-bonded pairs with a small number of
neighbors will on average have a stronger H-bond than bonded pairs with a
greater number of neighbors \cite{matysiak2011dissecting}. Furthermore, the
fraction of H-bonded pairs in interfacial regions of apolar moieties is lower
than in the bulk liquid, and the bonded pairs that do exist in these regions
tend to have fewer neighbors and stronger bonds than the average H-bonded pair
in the bulk.  The interpretation is that density fluctuations that create
cavities select against weak H-bonds, leaving only the stronger bonds to
survive. Thus, the interfacial region experiences less H-bonding on the whole
than equivalent volumes in the bulk, but maintains on average stronger hydrogen
bonds.

A plausible analogy in the Jagla liquid to H-bonding in water is the
interaction of particle pairs at the potential minimum distance, $r_1$.  As
temperature is lowered, the liquid prefers to adopt configurations that
maximize the number of particle pairs near a separation of $r_1$, which in the
limit of the crystal is an hcp lattice \cite{xu2006thermodynamics}.  This is
analogous to water maximizing the number of H-bonded pairs at low temperatures
by adopting a tetrahedral network structure, and thus the Jagla pair
interactions near $r_1$ become analogous to water's H-bond.  Under this view,
density fluctuations in the Jagla liquid disrupt weakly interacting Jagla
particles and leave a solvation shell that consists of fewer pair interactions
near $r_1$.  The fraction of ``broken'' interactions at $r_1$ in the solvation
shell would increase faster with temperature than the same quantity in the
bulk. Future work entailing a detailed analysis of solvation shell structure
will be needed to demonstrate if this hypothesis is correct.

The LJ liquid demonstrates enthalpic and entropic behaviors in sharp contrast
to those of water and the Jagla liquid.  The entropy is strictly positive for
all cavities of size $R>0.5\sigma_{LJ}$ in the LJ liquid, and the heat capacity
increment is negligible.  This latter phenomenon is consistent with the
argument for the temperature dependence of the relative fraction of broken
H-bonds in solvation shell water compared to bulk water---\emph{i.e.}, the
absence of a second energy scale in the LJ liquid precludes a
temperature-dependent enthalpy of cavity formation analogous to that of
water. This implies that the fundamental commonality between water and Jagla
fluids is the presence of two energy scales, each coupled to a different length
scale, so that low density, open structures are increasingly favored for
decreasing temperature, the feature absent in simple liquids. In the Jagla
model, the second energy and length scale is set by the ratios describing the
soft ramp, $\varepsilon_1/\varepsilon_2$ and $r_1/r_0$ , while in water, these
are determined by characteristics of the H-bonded and non-H-bonded states.

\subsection{The Length Scale Crossover}

As seen in Fig.~\ref{fig:mu_ex_per_a}, the chemical potential decreases with
temperature along the coexistence curve for all cavity sizes considered in the
LJ liquid.  However, in water and the Jagla liquid, the chemical potential
increases with increasing temperature for solvent-sized cavities and decreases
with temperature for larger cavities.  Qualitatively, the temperature
dependence of the solvation free energy is identical in the Jagla liquid and
water.
 
\begin{figure}[htbp]
  \centering
  \includegraphics[width=0.9\columnwidth]{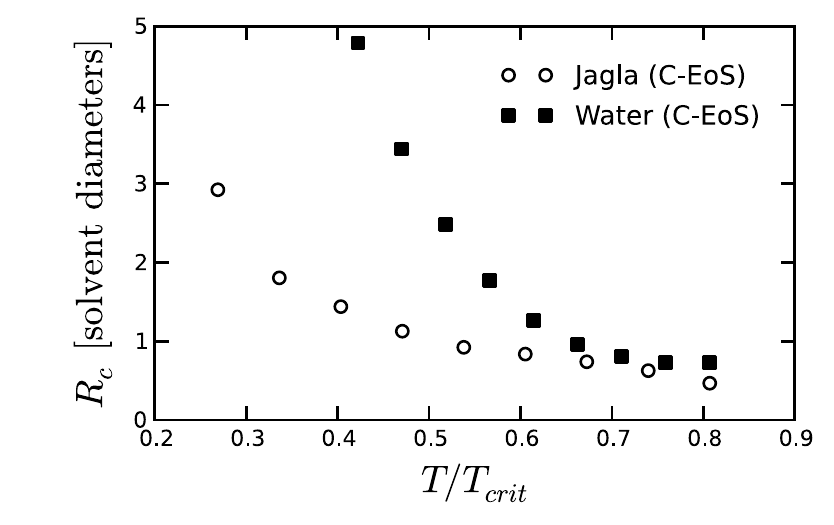}
  \caption{Entropic sign crossover lengths for cavity solutes in the Jagla
    liquid and water as predicted by the cavity equation of state.  The values
    are plotted as a function of temperature reduced by $T_{crit}$, the
    liquid-vapor critical point.  Points indicate the cavity radius, in units
    of solvent diameters, at which the solvation entropy changes sign from
    negative to positive.  Entropic crossovers for cavities in the LJ liquid
    also occur, but at cavity radii less than 0.5 solvent diameters for all
    states on the saturation curve (not shown).}
  \label{fig:entropic_crossovers}
\end{figure}

An important consequence of the similarities between the temperature-dependence
of the solvation free energies in the Jagla liquid and water is that the
water-like characteristic of negative solvation entropy for small cavities is
observed in the Jagla liquid (Fig.~\ref{fig:s_and_h_vs_R}).  As the cavity size
increases from $R=0.5\sigma_{JG}$, the curves along each saturation state first
decrease, then pass through a minimum before increasing monotonically for
larger cavities.  For cavities large enough that $s_c^x|_\sigma>0$, the
solvation shell is more disordered, and for sufficiently large cavities a
dewetting transition will occur.  This ``entropic crossover'' from negative to
positive solvation entropy may therefore be viewed as a measure of the length
scale at which interface formation begins to dominate the solvation free
energy.  In this view, the crossover for the LJ liquid occurs at cavity sizes
less than $\sigma_{LJ}$ in diameter for all saturation states, which is smaller
than the smallest cavities explicitly studied here.  In water and the Jagla
liquid however, the entropic crossover distance grows many times larger than
the solvent diameter as temperature is decreased, as shown in
Fig. \ref{fig:entropic_crossovers}.  Although the entropic crossover is similar
in the Jagla liquid and water, the crossover in water occurs at larger sizes
relative to the solvent diameter.

\subsection{The Thermodynamic Stability of Solvophobic Aggregates}

To explore the implications of the interplay between temperature and length
scale dependence of solvation free energies, we examine a simple picture of
solvophobic aggregation that combines ideas from Chandler
\cite{chandler2005interfaces} and Rajamani \emph{et al.}
\cite{rajamani2005hydrophobic}. Consider a solvophobic aggregate composed of
$n$ identical hard sphere particles with cavity radius $r$ such that the total
volume of the aggregate is $V = nv/\eta $, where $v$ is the volume of a single
constituent hard sphere particle and $\eta$ is the packing fraction of the
spheres. If the aggregate is treated as a large spherical volume of radius $R$,
then the aggregation Gibbs energy may be modeled as

\begin{equation}
  \label{eq:dGagg_noscale}
  \Delta G = \mu_{R} - n\mu_r,
\end{equation}

where $\mu_{R}$ is the aggregate's chemical potential and $\mu_r$ is the
chemical potential of a single constituent solvophobe at infinite dilution. The
relationship between the number of hard spheres comprising the aggregate and
its radius, $R$, is $n=4\pi \eta R^3/3v$. Combining the expressions for $n$ and
$\Delta G$ and dividing by the aggregate surface area, we have

\begin{align}
  \label{eq:G_agg}
  \Delta G(R) / 4\pi R^2 &= \mu_{R}(R)/4\pi R^2 - \mu_r \eta R/3v .
\end{align}

\begin{figure}[htbp]
  \centering
  \includegraphics[width=0.9\columnwidth]{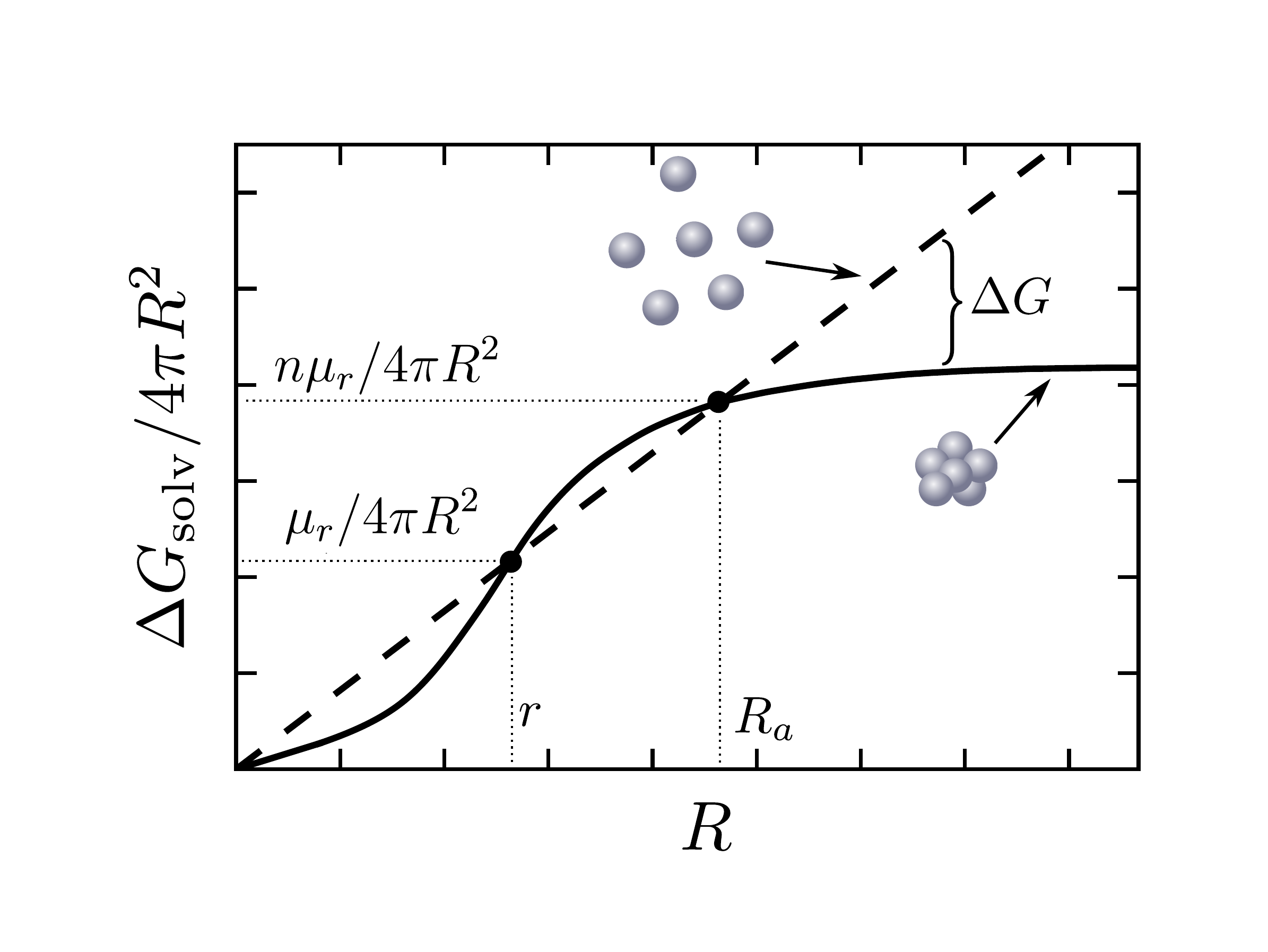}
  \caption{Solvation free energy scaled by the surface area versus aggregate
    radius.  The solid line correspond to the solvation free energy per unit
    surface area of a cavity of size $R$, which is used to model an aggregate
    of $n$ smaller hard spheres of size $r$ (see text). The dashed line
    represent the solvation free energy per unit surface area for $n$
    constituent spheres fully dispersed in solution.  Only aggregates larger
    than the aggregation radius, $R_a$, are thermodynamically stable.}
  \label{fig:aggregate_DG}
\end{figure}

For increasing $R$, the first term on the RHS of Eq. (\ref{eq:G_agg}) becomes
approximately constant and equal to the interfacial free energy per unit area
\cite{huang2001scaling}.  The second term is a linear function of the aggregate
radius. The radius at which the RHS vanishes is the aggregation radius,
$R_a$---aggregates of size larger than $R_a$ are thermodynamically stable
within this model free energy.  These concepts are shown pictorially in
Fig. \ref{fig:aggregate_DG}.

\begin{figure}[htbp]
  \centering
  \includegraphics[width=0.8\columnwidth]{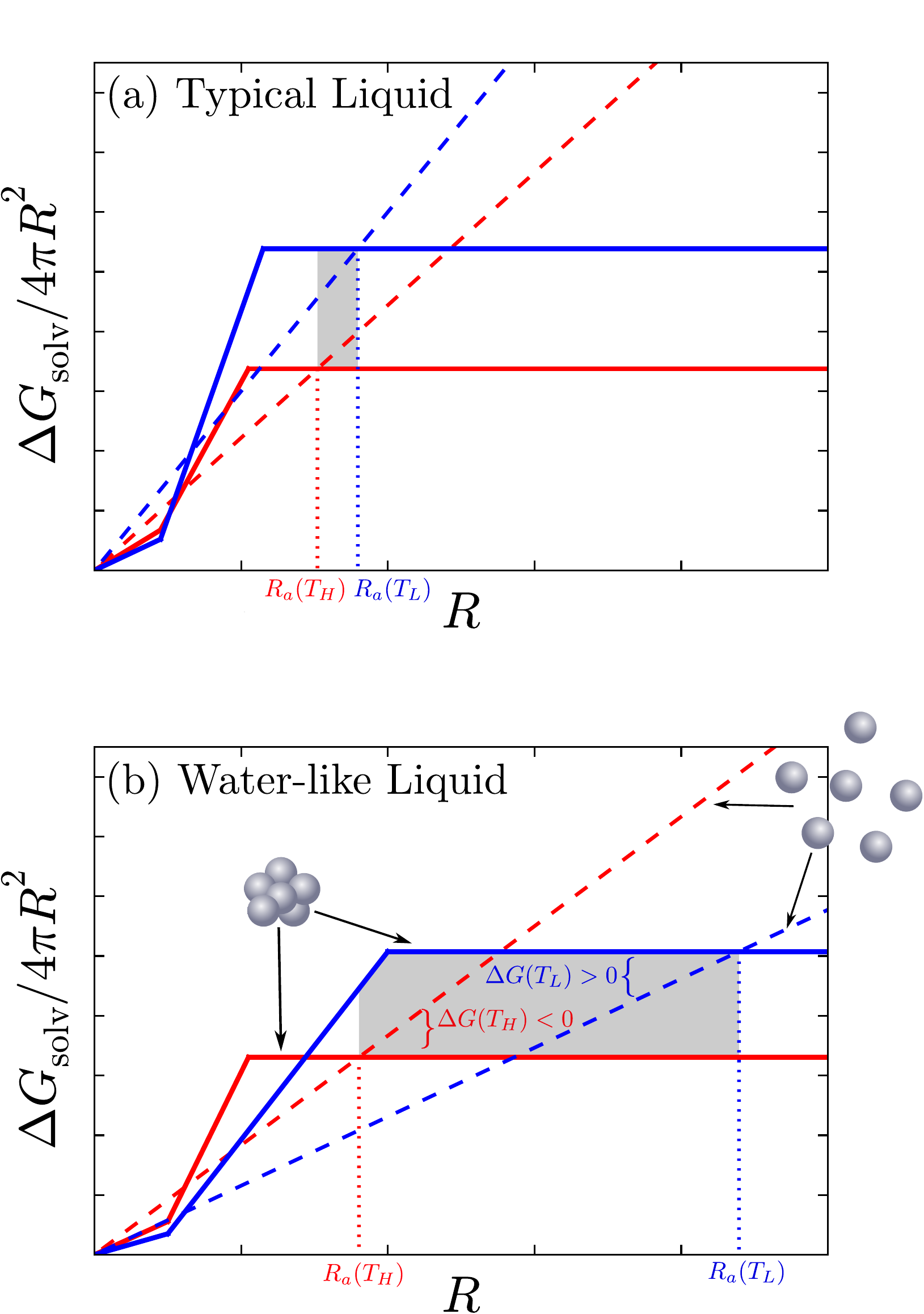}
  \caption{Qualitative depiction of solvation free energy per surface area of
    large solvophobic aggregates and dispersed small solutes in (a) typical and
    (b) water-like solvents.  Red and blue correspond to warm ($T_H$) and cold
    temperatures ($T_L$), respectively.  The shaded region highlights the
    aggregate size range where cooling from $T_H$ to $T_L$ destabilizes the
    aggregate.  The sloped line which here depicts the rise from very small
    solute to large radius behavior is used to emphasize that the shape of this
    molecular scale transition region is represented only generically in this
    figure.}
  \label{fig:water_vs_typical}
\end{figure}

\begin{figure}[htbp]
  \centering
  \includegraphics[width=0.9\columnwidth]{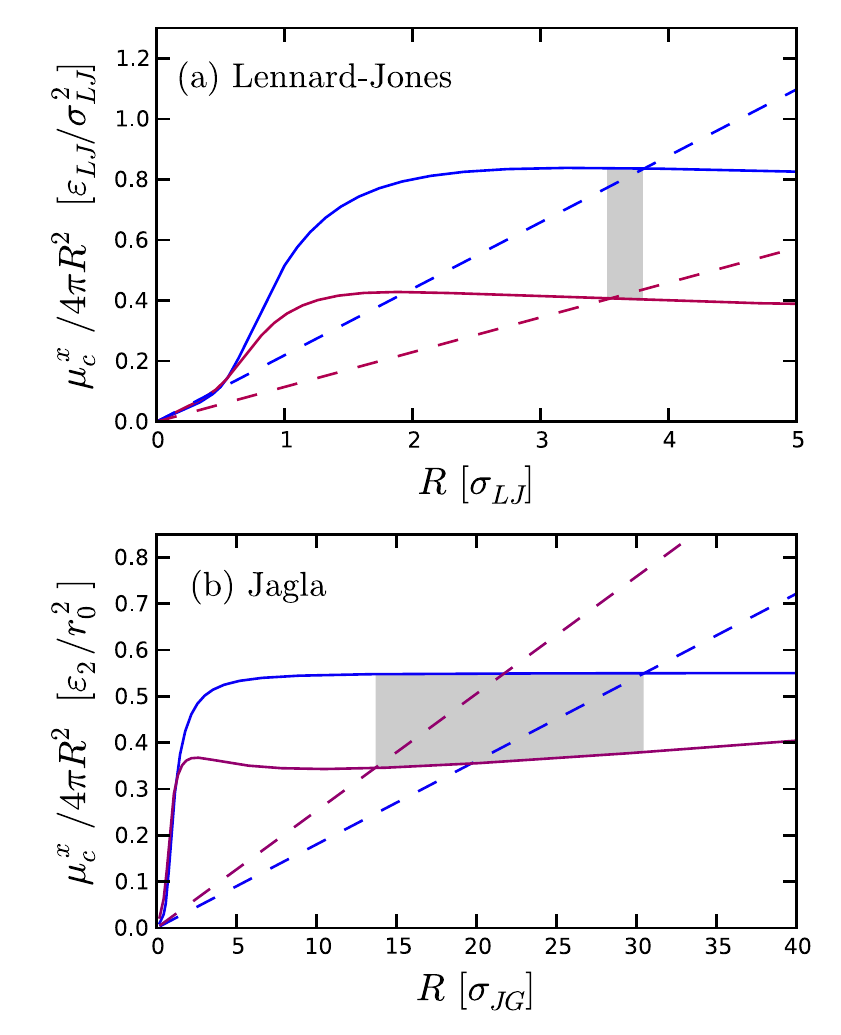}
  \caption{The specific case of Fig. \ref{fig:water_vs_typical} for the
    temperature dependence of solvophobic solvation free energies in (a) the LJ
    liquid for $T=0.65$ (blue) and $T=0.95$ [$\varepsilon_{LJ}/k_B$] (red) and
    (b) the Jagla liquid for $T=0.4$ (blue) and $T=1.0$ [$\varepsilon_{2}/k_B$]
    (red).  The constituent solvophobes are equivalent in size to the solvent
    diameter and the aggregate packing fraction is taken equivalent to the
    solvent packing fraction.  Both liquids have a range of cavity sizes
    (shaded region) where cooling from the warm temperature (red lines) to the
    cool temperature (blue lines) destabilizes the aggregate (solid lines)
    relative to the dispersed spheres (dashed lines).  The size range in the
    Jagla liquid is far more pronounced, however (note the order of magnitude
    difference in the abscissa scales).}
  \label{fig:jagla_and_lj_aggregate_cool}
\end{figure}

We now consider the process of cooling the aggregate from a warm temperature,
$T_H$, to a lower temperature, $T_L$, and in particular, the effect that this
process has on the thermodynamic stability of the aggregate. A qualitative
picture of the dependence of the aggregation radius, $R_a$, on temperature for
a water-like and a reference LJ-like fluid is shown in
Fig. \ref{fig:water_vs_typical}.  The differences in crossover behavior arise
due to the fact that for small solutes in water-like solvents, increasing the
temperature decreases the solubility.  This has two effects: the first is that
the crossover length scale is more sensitive to temperature, and the second is
that the slope of the dispersed solvophobes line for high temperature is
greater than the corresponding line at low temperature.  These effects combine
to produce a range of aggregate sizes that are thermodynamically stable at
$T_H$ but become unstable upon cooling a to $T_L$.  It is interesting that such
a region also appears in a typical LJ-like liquid.  However, the crossover
length scale in LJ-like liquids is less sensitive to temperature and the slope
of the dispersed solvophobes line is greater at lower temperatures, causing the
region of destabilization to dramatically shrink or altogether disappear.
Fig. \ref{fig:jagla_and_lj_aggregate_cool} shows quantitative measures of the
dissociation size range in the LJ and Jagla liquids for cavities equivalent to
the solvent size and aggregate packing fractions equivalent to the solvent
packing fraction.  The dissociation region in the Jagla liquid is orders of
magnitude larger than that in the LJ liquid.

In general, the range of the destabilization region is extended by cooling to
lower temperatures or by composing aggregates of smaller constituent particles.
A prediction made by this model is the possibility of cold-induced dissociation
of solvophobic aggregates in LJ-like solvents.  Aggregates composed of
sufficiently small cavity solutes will in fact, in this model, have a range of
sizes for which cooling will destabilize the aggregate and induce its
decomposition.  It would indeed be striking if such a limit were faithfully
captured by this thought experiment in spite of its overall simplicity.

\section{Conclusions}
\label{sec:conclusions}

The results of exhaustive MC simulations of cavity formation along the
saturation curves of the LJ liquid and the Jagla liquid were presented.  The
temperature-dependence of the solvation thermodynamics of cavities ranging from
one-half to six times the solvent particle size were compared between the two
simple liquids and to predictions for cavity formation in water given by a
cavity equation of state (C-EoS).  The comparisons between the Jagla liquid,
water, and the simple liquid (LJ) serve to illuminate the features of
hydrophobic hydration that are unique to water.

The Jagla liquid demonstrates water-like behavior in its resistance to
dewetting of large cavity surfaces.  In the presence of the largest cavity
sizes considered (six solvent diameters), the LJ liquid showed a dewetting
transition at all thermodynamic states on the saturation curve, whereas the
Jagla liquid resists dewetting at low temperature saturation states.

The Jagla liquid is also water-like in its enthalpic and entropic behavior in
the sense that the solvation entropy of small cavities is negative and the heat
capacity increment is positive. The LJ liquid on the other hand manifests a
strictly positive entropy for all cavities larger than half the solvent size
and shows a negligible heat capacity increment.

From our analysis, we infer the important result that it is the existence of a
second energy scale in the Jagla liquid and in water, compared to a simple
liquid, that energetically favors the creation of void space at low
temperatures, that gives rise to the anomalous liquid state properties as well
as solvation behavior.  Of course, the ability of the fluid to access the low
energy structures with only modest expansion implies that the particular length
scales involved are closely coupled to this observation \cite{yan2006family}.


We have demonstrated that the scaling and temperature dependence of the
solvation free energies of cavity solutes in Jagla liquid is qualitatively
similar to that of water.  Both liquids have negative solvation entropies for
small cavities that cross over to positive with increasing cavity size.  These
crossovers for the Jagla liquid occur at a shorter length scale relative to the
solvent size than those of water.

Combining ideas from Chandler \cite{chandler2005interfaces} and Rajamani
\emph{et al.} \cite{rajamani2005hydrophobic}, a simple model for aggregate
dissociation was introduced by modeling an aggregate as a single large hard
sphere with a volume equal to the sum of the volumes of the constituent spheres
divided by a packing fraction.  The consequences of the differing size scaling
and temperature dependence of solvation free energy for the aggregate compared
to the dispersed constituent spheres is clearly demonstrated in the context of
this simple model for aggregation.  In particular, it was shown that
cold-induced dissociation will occur for aggregates composed of sufficiently
small spheres in water-like liquids.  The degree to which such behavior is
accurately described by the simple model is of interest for further
investigations, as is the detailed examination of other two-scale liquids
containing both a hard and soft core component.

\begin{acknowledgments}
  The authors are grateful to Henry S. Ashbaugh for providing us with the
  numerical results of his calculations on the Lennard-Jones fluid. This
  project was supported by the National Science Foundation (CHE-0910615) with
  additional support from the R. A. Welch Foundation (F-0019). PGD gratefully
  acknowledges the support of the National Science Foundation (CHE-1213343).
  Computations were performed at the Texas Advanced Computing Center.
\end{acknowledgments}


\bibliographystyle{apsrev}
\bibliography{TPL_dep_refs}

\begin{thebibliography}{3}
\expandafter\ifx\csname natexlab\endcsname\relax\def\natexlab#1{#1}\fi
\expandafter\ifx\csname bibnamefont\endcsname\relax
  \def\bibnamefont#1{#1}\fi
\expandafter\ifx\csname bibfnamefont\endcsname\relax
  \def\bibfnamefont#1{#1}\fi
\expandafter\ifx\csname citenamefont\endcsname\relax
  \def\citenamefont#1{#1}\fi
\expandafter\ifx\csname url\endcsname\relax
  \def\url#1{\texttt{#1}}\fi
\expandafter\ifx\csname urlprefix\endcsname\relax\def\urlprefix{URL }\fi
\providecommand{\bibinfo}[2]{#2}
\providecommand{\eprint}[2][]{\url{#2}}

\bibitem[{\citenamefont{Ashbaugh}(2009)}]{ashbaugh2009blowing}
\bibinfo{author}{\bibfnamefont{H.~S.} \bibnamefont{Ashbaugh}},
  \bibinfo{journal}{The Journal of Chemical Physics}
  \textbf{\bibinfo{volume}{130}}, \bibinfo{eid}{204517} (\bibinfo{year}{2009}).

\bibitem[{\citenamefont{Guissani and Guillot}(1993)}]{guissani1993coexistence}
\bibinfo{author}{\bibfnamefont{Y.}~\bibnamefont{Guissani}} \bibnamefont{and}
  \bibinfo{author}{\bibfnamefont{B.}~\bibnamefont{Guillot}},
  \bibinfo{journal}{The Journal of Chemical Physics}
  \textbf{\bibinfo{volume}{98}}, \bibinfo{pages}{8221} (\bibinfo{year}{1993}).

\bibitem[{\citenamefont{Ben-Amotz}(2005)}]{ben2005global}
\bibinfo{author}{\bibfnamefont{D.}~\bibnamefont{Ben-Amotz}},
  \bibinfo{journal}{The Journal of Chemical Physics}
  \textbf{\bibinfo{volume}{123}}, \bibinfo{eid}{184504} (\bibinfo{year}{2005}).

\end{thebibliography}


\begin{thebibliography}{47}
\expandafter\ifx\csname natexlab\endcsname\relax\def\natexlab#1{#1}\fi
\expandafter\ifx\csname bibnamefont\endcsname\relax
  \def\bibnamefont#1{#1}\fi
\expandafter\ifx\csname bibfnamefont\endcsname\relax
  \def\bibfnamefont#1{#1}\fi
\expandafter\ifx\csname citenamefont\endcsname\relax
  \def\citenamefont#1{#1}\fi
\expandafter\ifx\csname url\endcsname\relax
  \def\url#1{\texttt{#1}}\fi
\expandafter\ifx\csname urlprefix\endcsname\relax\def\urlprefix{URL }\fi
\providecommand{\bibinfo}[2]{#2}
\providecommand{\eprint}[2][]{\url{#2}}

\bibitem[{\citenamefont{Chandler}(2005)}]{chandler2005interfaces}
\bibinfo{author}{\bibfnamefont{D.}~\bibnamefont{Chandler}},
  \bibinfo{journal}{Nature} \textbf{\bibinfo{volume}{437}},
  \bibinfo{pages}{640} (\bibinfo{year}{2005}).

\bibitem[{\citenamefont{Lum et~al.}(1999)\citenamefont{Lum, Chandler, and
  Weeks}}]{lum1999hydrophobicity}
\bibinfo{author}{\bibfnamefont{K.}~\bibnamefont{Lum}},
  \bibinfo{author}{\bibfnamefont{D.}~\bibnamefont{Chandler}}, \bibnamefont{and}
  \bibinfo{author}{\bibfnamefont{J.~D.} \bibnamefont{Weeks}},
  \bibinfo{journal}{The Journal of Physical Chemistry B}
  \textbf{\bibinfo{volume}{103}}, \bibinfo{pages}{4570} (\bibinfo{year}{1999}).

\bibitem[{\citenamefont{Huang et~al.}(2001)\citenamefont{Huang, Geissler, and
  Chandler}}]{huang2001scaling}
\bibinfo{author}{\bibfnamefont{D.~M.} \bibnamefont{Huang}},
  \bibinfo{author}{\bibfnamefont{P.~L.} \bibnamefont{Geissler}},
  \bibnamefont{and} \bibinfo{author}{\bibfnamefont{D.}~\bibnamefont{Chandler}},
  \bibinfo{journal}{The Journal of Physical Chemistry B}
  \textbf{\bibinfo{volume}{105}}, \bibinfo{pages}{6704} (\bibinfo{year}{2001}).

\bibitem[{\citenamefont{Huang and Chandler}(2002)}]{huang2002hydrophobic}
\bibinfo{author}{\bibfnamefont{D.~M.} \bibnamefont{Huang}} \bibnamefont{and}
  \bibinfo{author}{\bibfnamefont{D.}~\bibnamefont{Chandler}},
  \bibinfo{journal}{The Journal of Physical Chemistry B}
  \textbf{\bibinfo{volume}{106}}, \bibinfo{pages}{2047} (\bibinfo{year}{2002}).

\bibitem[{\citenamefont{Huang and Chandler}(2000)}]{huang2000cavity}
\bibinfo{author}{\bibfnamefont{D.~M.} \bibnamefont{Huang}} \bibnamefont{and}
  \bibinfo{author}{\bibfnamefont{D.}~\bibnamefont{Chandler}},
  \bibinfo{journal}{Phys. Rev. E} \textbf{\bibinfo{volume}{61}},
  \bibinfo{pages}{1501} (\bibinfo{year}{2000}).

\bibitem[{\citenamefont{Rajamani et~al.}(2005)\citenamefont{Rajamani, Truskett,
  and Garde}}]{rajamani2005hydrophobic}
\bibinfo{author}{\bibfnamefont{S.}~\bibnamefont{Rajamani}},
  \bibinfo{author}{\bibfnamefont{T.~M.} \bibnamefont{Truskett}},
  \bibnamefont{and} \bibinfo{author}{\bibfnamefont{S.}~\bibnamefont{Garde}},
  \bibinfo{journal}{Proceedings of the National Academy of Sciences}
  \textbf{\bibinfo{volume}{102}}, \bibinfo{pages}{9475} (\bibinfo{year}{2005}).

\bibitem[{\citenamefont{Patel et~al.}(2010)\citenamefont{Patel, Varilly, and
  Chandler}}]{patel2010fluctuations}
\bibinfo{author}{\bibfnamefont{A.~J.} \bibnamefont{Patel}},
  \bibinfo{author}{\bibfnamefont{P.}~\bibnamefont{Varilly}}, \bibnamefont{and}
  \bibinfo{author}{\bibfnamefont{D.}~\bibnamefont{Chandler}},
  \bibinfo{journal}{The Journal of Physical Chemistry B}
  \textbf{\bibinfo{volume}{114}}, \bibinfo{pages}{1632} (\bibinfo{year}{2010}).

\bibitem[{\citenamefont{Yan et~al.}(2005)\citenamefont{Yan, Buldyrev,
  Giovambattista, and Stanley}}]{yan2005structural}
\bibinfo{author}{\bibfnamefont{Z.}~\bibnamefont{Yan}},
  \bibinfo{author}{\bibfnamefont{S.~V.} \bibnamefont{Buldyrev}},
  \bibinfo{author}{\bibfnamefont{N.}~\bibnamefont{Giovambattista}},
  \bibnamefont{and} \bibinfo{author}{\bibfnamefont{H.~E.}
  \bibnamefont{Stanley}}, \bibinfo{journal}{Phys. Rev. Lett.}
  \textbf{\bibinfo{volume}{95}}, \bibinfo{pages}{130604}
  (\bibinfo{year}{2005}).

\bibitem[{\citenamefont{Yan et~al.}(2006)\citenamefont{Yan, Buldyrev,
  Giovambattista, Debenedetti, and Stanley}}]{yan2006family}
\bibinfo{author}{\bibfnamefont{Z.}~\bibnamefont{Yan}},
  \bibinfo{author}{\bibfnamefont{S.~V.} \bibnamefont{Buldyrev}},
  \bibinfo{author}{\bibfnamefont{N.}~\bibnamefont{Giovambattista}},
  \bibinfo{author}{\bibfnamefont{P.~G.} \bibnamefont{Debenedetti}},
  \bibnamefont{and} \bibinfo{author}{\bibfnamefont{H.~E.}
  \bibnamefont{Stanley}}, \bibinfo{journal}{Phys. Rev. E}
  \textbf{\bibinfo{volume}{73}}, \bibinfo{pages}{051204}
  (\bibinfo{year}{2006}).

\bibitem[{\citenamefont{Buldyrev et~al.}(2009)\citenamefont{Buldyrev, Malescio,
  Angell, Giovambattista, Prestipino, Saija, Stanley, and
  Xu}}]{buldyrev_unusual_2009}
\bibinfo{author}{\bibfnamefont{S.~V.} \bibnamefont{Buldyrev}},
  \bibinfo{author}{\bibfnamefont{G.}~\bibnamefont{Malescio}},
  \bibinfo{author}{\bibfnamefont{C.~A.} \bibnamefont{Angell}},
  \bibinfo{author}{\bibfnamefont{N.}~\bibnamefont{Giovambattista}},
  \bibinfo{author}{\bibfnamefont{S.}~\bibnamefont{Prestipino}},
  \bibinfo{author}{\bibfnamefont{F.}~\bibnamefont{Saija}},
  \bibinfo{author}{\bibfnamefont{H.~E.} \bibnamefont{Stanley}},
  \bibnamefont{and} \bibinfo{author}{\bibfnamefont{L.}~\bibnamefont{Xu}},
  \bibinfo{journal}{Journal of Physics: Condensed Matter}
  \textbf{\bibinfo{volume}{21}}, \bibinfo{pages}{504106}
  (\bibinfo{year}{2009}).

\bibitem[{\citenamefont{Jagla}(1998)}]{jagla1998phase}
\bibinfo{author}{\bibfnamefont{E.~A.} \bibnamefont{Jagla}},
  \bibinfo{journal}{Phys. Rev. E} \textbf{\bibinfo{volume}{58}},
  \bibinfo{pages}{1478} (\bibinfo{year}{1998}).

\bibitem[{\citenamefont{Jagla}(1999)}]{jagla1999core}
\bibinfo{author}{\bibfnamefont{E.~A.} \bibnamefont{Jagla}},
  \bibinfo{journal}{The Journal of Chemical Physics}
  \textbf{\bibinfo{volume}{111}}, \bibinfo{pages}{8980} (\bibinfo{year}{1999}).

\bibitem[{\citenamefont{Buldyrev et~al.}(2007)\citenamefont{Buldyrev, Kumar,
  Debenedetti, Rossky, and Stanley}}]{buldyrev2007water}
\bibinfo{author}{\bibfnamefont{S.~V.} \bibnamefont{Buldyrev}},
  \bibinfo{author}{\bibfnamefont{P.}~\bibnamefont{Kumar}},
  \bibinfo{author}{\bibfnamefont{P.~G.} \bibnamefont{Debenedetti}},
  \bibinfo{author}{\bibfnamefont{P.~J.} \bibnamefont{Rossky}},
  \bibnamefont{and} \bibinfo{author}{\bibfnamefont{H.~E.}
  \bibnamefont{Stanley}}, \bibinfo{journal}{Proceedings of the National Academy
  of Sciences} \textbf{\bibinfo{volume}{104}}, \bibinfo{pages}{20177}
  (\bibinfo{year}{2007}).

\bibitem[{\citenamefont{Buldyrev et~al.}(2010)\citenamefont{Buldyrev, Kumar,
  Sastry, Stanley, and Weiner}}]{buldyrev_hydrophobic_2010}
\bibinfo{author}{\bibfnamefont{S.~V.} \bibnamefont{Buldyrev}},
  \bibinfo{author}{\bibfnamefont{P.}~\bibnamefont{Kumar}},
  \bibinfo{author}{\bibfnamefont{S.}~\bibnamefont{Sastry}},
  \bibinfo{author}{\bibfnamefont{H.~E.} \bibnamefont{Stanley}},
  \bibnamefont{and} \bibinfo{author}{\bibfnamefont{S.}~\bibnamefont{Weiner}},
  \bibinfo{journal}{Journal of Physics: Condensed Matter}
  \textbf{\bibinfo{volume}{22}}, \bibinfo{pages}{284109}
  (\bibinfo{year}{2010}).

\bibitem[{\citenamefont{Maiti et~al.}(2012)\citenamefont{Maiti, Weiner,
  Buldyrev, Stanley, and Sastry}}]{maiti_potential_2012}
\bibinfo{author}{\bibfnamefont{M.}~\bibnamefont{Maiti}},
  \bibinfo{author}{\bibfnamefont{S.}~\bibnamefont{Weiner}},
  \bibinfo{author}{\bibfnamefont{S.~V.} \bibnamefont{Buldyrev}},
  \bibinfo{author}{\bibfnamefont{H.~E.} \bibnamefont{Stanley}},
  \bibnamefont{and} \bibinfo{author}{\bibfnamefont{S.}~\bibnamefont{Sastry}},
  \bibinfo{journal}{The Journal of Chemical Physics}
  \textbf{\bibinfo{volume}{136}}, \bibinfo{eid}{044512} (\bibinfo{year}{2012}).

\bibitem[{\citenamefont{Molinero and Moore}(2009)}]{molinero2009intermediate}
\bibinfo{author}{\bibfnamefont{V.}~\bibnamefont{Molinero}} \bibnamefont{and}
  \bibinfo{author}{\bibfnamefont{E.~B.} \bibnamefont{Moore}},
  \bibinfo{journal}{The Journal of Physical Chemistry B}
  \textbf{\bibinfo{volume}{113}}, \bibinfo{pages}{4008} (\bibinfo{year}{2009}).

\bibitem[{\citenamefont{Moore and Molinero}(2010)}]{moore2010ice}
\bibinfo{author}{\bibfnamefont{E.~B.} \bibnamefont{Moore}} \bibnamefont{and}
  \bibinfo{author}{\bibfnamefont{V.}~\bibnamefont{Molinero}},
  \bibinfo{journal}{The Journal of Chemical Physics}
  \textbf{\bibinfo{volume}{132}}, \bibinfo{pages}{244504}
  (\bibinfo{year}{2010}).

\bibitem[{\citenamefont{Moore and Molinero}(2011)}]{moore2011structural}
\bibinfo{author}{\bibfnamefont{E.~B.} \bibnamefont{Moore}} \bibnamefont{and}
  \bibinfo{author}{\bibfnamefont{V.}~\bibnamefont{Molinero}},
  \bibinfo{journal}{Nature} \textbf{\bibinfo{volume}{479}},
  \bibinfo{pages}{506} (\bibinfo{year}{2011}).

\bibitem[{\citenamefont{Widom}(1963)}]{widom1963some}
\bibinfo{author}{\bibfnamefont{B.}~\bibnamefont{Widom}}, \bibinfo{journal}{The
  Journal of Chemical Physics} \textbf{\bibinfo{volume}{39}},
  \bibinfo{pages}{2808} (\bibinfo{year}{1963}).

\bibitem[{\citenamefont{Widom}(1982)}]{widom1982potential}
\bibinfo{author}{\bibfnamefont{B.}~\bibnamefont{Widom}}, \bibinfo{journal}{The
  Journal of Physical Chemistry} \textbf{\bibinfo{volume}{86}},
  \bibinfo{pages}{869} (\bibinfo{year}{1982}).

\bibitem[{\citenamefont{Frenkel and Smit}(2001)}]{frenkel2002understanding}
\bibinfo{author}{\bibfnamefont{D.}~\bibnamefont{Frenkel}} \bibnamefont{and}
  \bibinfo{author}{\bibfnamefont{B.}~\bibnamefont{Smit}},
  \emph{\bibinfo{title}{{Understanding Molecular Simulation: From Algorithms to
  Applications}}} (\bibinfo{publisher}{Academic Press}, \bibinfo{year}{2001}).

\bibitem[{\citenamefont{Kofke and Cummings}(1997)}]{kofke1997quantitative}
\bibinfo{author}{\bibfnamefont{D.~A.} \bibnamefont{Kofke}} \bibnamefont{and}
  \bibinfo{author}{\bibfnamefont{P.~T.} \bibnamefont{Cummings}},
  \bibinfo{journal}{Molecular Physics} \textbf{\bibinfo{volume}{92}},
  \bibinfo{pages}{973} (\bibinfo{year}{1997}).

\bibitem[{\citenamefont{Ashbaugh and Pratt}(2006)}]{ashbaugh2006colloquium}
\bibinfo{author}{\bibfnamefont{H.~S.} \bibnamefont{Ashbaugh}} \bibnamefont{and}
  \bibinfo{author}{\bibfnamefont{L.~R.} \bibnamefont{Pratt}},
  \bibinfo{journal}{Rev. Mod. Phys.} \textbf{\bibinfo{volume}{78}},
  \bibinfo{pages}{159} (\bibinfo{year}{2006}).

\bibitem[{\citenamefont{Ashbaugh and Pratt}(2007)}]{ashbaugh2007contrasting}
\bibinfo{author}{\bibfnamefont{H.~S.} \bibnamefont{Ashbaugh}} \bibnamefont{and}
  \bibinfo{author}{\bibfnamefont{L.~R.} \bibnamefont{Pratt}},
  \bibinfo{journal}{The Journal of Physical Chemistry B}
  \textbf{\bibinfo{volume}{111}}, \bibinfo{pages}{9330} (\bibinfo{year}{2007}).

\bibitem[{\citenamefont{Ashbaugh}(2009)}]{ashbaugh2009blowing}
\bibinfo{author}{\bibfnamefont{H.~S.} \bibnamefont{Ashbaugh}},
  \bibinfo{journal}{The Journal of Chemical Physics}
  \textbf{\bibinfo{volume}{130}}, \bibinfo{eid}{204517} (\bibinfo{year}{2009}).

\bibitem[{\citenamefont{Reiss et~al.}(1959)\citenamefont{Reiss, Frisch, and
  Lebowitz}}]{reiss1959statistical}
\bibinfo{author}{\bibfnamefont{H.}~\bibnamefont{Reiss}},
  \bibinfo{author}{\bibfnamefont{H.~L.} \bibnamefont{Frisch}},
  \bibnamefont{and} \bibinfo{author}{\bibfnamefont{J.~L.}
  \bibnamefont{Lebowitz}}, \bibinfo{journal}{The Journal of Chemical Physics}
  \textbf{\bibinfo{volume}{31}}, \bibinfo{pages}{369} (\bibinfo{year}{1959}).

\bibitem[{\citenamefont{Stillinger}(1973)}]{stillinger1973structure}
\bibinfo{author}{\bibfnamefont{F.~H.} \bibnamefont{Stillinger}},
  \bibinfo{journal}{Journal of Solution Chemistry}
  \textbf{\bibinfo{volume}{2}}, \bibinfo{pages}{141} (\bibinfo{year}{1973}).

\bibitem[{\citenamefont{Tully-Smith and Reiss}(1970)}]{tully1970further}
\bibinfo{author}{\bibfnamefont{D.~M.} \bibnamefont{Tully-Smith}}
  \bibnamefont{and} \bibinfo{author}{\bibfnamefont{H.}~\bibnamefont{Reiss}},
  \bibinfo{journal}{The Journal of Chemical Physics}
  \textbf{\bibinfo{volume}{53}}, \bibinfo{pages}{4015} (\bibinfo{year}{1970}).

\bibitem[{\citenamefont{Stillinger and Cotter}(1971)}]{stillinger1971free}
\bibinfo{author}{\bibfnamefont{F.~H.} \bibnamefont{Stillinger}}
  \bibnamefont{and} \bibinfo{author}{\bibfnamefont{M.~A.}
  \bibnamefont{Cotter}}, \bibinfo{journal}{The Journal of Chemical Physics}
  \textbf{\bibinfo{volume}{55}}, \bibinfo{pages}{3449} (\bibinfo{year}{1971}).

\bibitem[{\citenamefont{Ben-Amotz}(2005)}]{ben2005global}
\bibinfo{author}{\bibfnamefont{D.}~\bibnamefont{Ben-Amotz}},
  \bibinfo{journal}{The Journal of Chemical Physics}
  \textbf{\bibinfo{volume}{123}}, \bibinfo{eid}{184504} (\bibinfo{year}{2005}).

\bibitem[{\citenamefont{Errington and
  Debenedetti}(2001)}]{errington2001relationship}
\bibinfo{author}{\bibfnamefont{J.~R.} \bibnamefont{Errington}}
  \bibnamefont{and} \bibinfo{author}{\bibfnamefont{P.~G.}
  \bibnamefont{Debenedetti}}, \bibinfo{journal}{Nature}
  \textbf{\bibinfo{volume}{409}}, \bibinfo{pages}{318} (\bibinfo{year}{2001}).

\bibitem[{\citenamefont{Xu et~al.}(2006)\citenamefont{Xu, Buldyrev, Angell, and
  Stanley}}]{xu2006thermodynamics}
\bibinfo{author}{\bibfnamefont{L.}~\bibnamefont{Xu}},
  \bibinfo{author}{\bibfnamefont{S.~V.} \bibnamefont{Buldyrev}},
  \bibinfo{author}{\bibfnamefont{C.~A.} \bibnamefont{Angell}},
  \bibnamefont{and} \bibinfo{author}{\bibfnamefont{H.~E.}
  \bibnamefont{Stanley}}, \bibinfo{journal}{Phys. Rev. E}
  \textbf{\bibinfo{volume}{74}}, \bibinfo{pages}{031108}
  (\bibinfo{year}{2006}).

\bibitem[{\citenamefont{Kirkwood and Buff}(1949)}]{kirkwood1949statistical}
\bibinfo{author}{\bibfnamefont{J.~G.} \bibnamefont{Kirkwood}} \bibnamefont{and}
  \bibinfo{author}{\bibfnamefont{F.~P.} \bibnamefont{Buff}},
  \bibinfo{journal}{The Journal of Chemical Physics}
  \textbf{\bibinfo{volume}{17}}, \bibinfo{pages}{338} (\bibinfo{year}{1949}).

\bibitem[{\citenamefont{Walton et~al.}(1983)\citenamefont{Walton, Tildesley,
  Rowlinson, and Henderson}}]{walton1983pressure}
\bibinfo{author}{\bibfnamefont{J.~P. R.~B.} \bibnamefont{Walton}},
  \bibinfo{author}{\bibfnamefont{D.~J.} \bibnamefont{Tildesley}},
  \bibinfo{author}{\bibfnamefont{J.~S.} \bibnamefont{Rowlinson}},
  \bibnamefont{and} \bibinfo{author}{\bibfnamefont{J.~R.}
  \bibnamefont{Henderson}}, \bibinfo{journal}{Molecular Physics}
  \textbf{\bibinfo{volume}{48}}, \bibinfo{pages}{1357} (\bibinfo{year}{1983}).

\bibitem[{\citenamefont{Berendsen et~al.}(1987)\citenamefont{Berendsen,
  Grigera, and Straatsma}}]{berendsen1987missing}
\bibinfo{author}{\bibfnamefont{H.~J.~C.} \bibnamefont{Berendsen}},
  \bibinfo{author}{\bibfnamefont{J.~R.} \bibnamefont{Grigera}},
  \bibnamefont{and} \bibinfo{author}{\bibfnamefont{T.~P.}
  \bibnamefont{Straatsma}}, \bibinfo{journal}{The Journal of Physical
  Chemistry} \textbf{\bibinfo{volume}{91}}, \bibinfo{pages}{6269}
  (\bibinfo{year}{1987}).

\bibitem[{\citenamefont{Berendsen et~al.}(1995)\citenamefont{Berendsen, Van
  Der~Spoel, and van Drunen}}]{berendsen1995gromacs}
\bibinfo{author}{\bibfnamefont{H.~J.~C.} \bibnamefont{Berendsen}},
  \bibinfo{author}{\bibfnamefont{D.}~\bibnamefont{Van Der~Spoel}},
  \bibnamefont{and} \bibinfo{author}{\bibfnamefont{R.}~\bibnamefont{van
  Drunen}}, \bibinfo{journal}{Computer Physics Communications}
  \textbf{\bibinfo{volume}{91}}, \bibinfo{pages}{43 } (\bibinfo{year}{1995}).

\bibitem[{\citenamefont{Van Der~Spoel et~al.}(2005)\citenamefont{Van Der~Spoel,
  Lindahl, Hess, Groenhof, Mark, and Berendsen}}]{van2005gromacs}
\bibinfo{author}{\bibfnamefont{D.}~\bibnamefont{Van Der~Spoel}},
  \bibinfo{author}{\bibfnamefont{E.}~\bibnamefont{Lindahl}},
  \bibinfo{author}{\bibfnamefont{B.}~\bibnamefont{Hess}},
  \bibinfo{author}{\bibfnamefont{G.}~\bibnamefont{Groenhof}},
  \bibinfo{author}{\bibfnamefont{A.~E.} \bibnamefont{Mark}}, \bibnamefont{and}
  \bibinfo{author}{\bibfnamefont{H.~J.~C.} \bibnamefont{Berendsen}},
  \bibinfo{journal}{Journal of Computational Chemistry}
  \textbf{\bibinfo{volume}{26}}, \bibinfo{pages}{1701} (\bibinfo{year}{2005}).

\bibitem[{\citenamefont{Miyamoto and Kollman}(1992)}]{miyamoto1992settle}
\bibinfo{author}{\bibfnamefont{S.}~\bibnamefont{Miyamoto}} \bibnamefont{and}
  \bibinfo{author}{\bibfnamefont{P.~A.} \bibnamefont{Kollman}},
  \bibinfo{journal}{Journal of Computational Chemistry}
  \textbf{\bibinfo{volume}{13}}, \bibinfo{pages}{952} (\bibinfo{year}{1992}).

\bibitem[{\citenamefont{Bussi et~al.}(2007)\citenamefont{Bussi, Donadio, and
  Parrinello}}]{bussi2007canonical}
\bibinfo{author}{\bibfnamefont{G.}~\bibnamefont{Bussi}},
  \bibinfo{author}{\bibfnamefont{D.}~\bibnamefont{Donadio}}, \bibnamefont{and}
  \bibinfo{author}{\bibfnamefont{M.}~\bibnamefont{Parrinello}},
  \bibinfo{journal}{Journal of Chemical Physics}
  \textbf{\bibinfo{volume}{126}}, \bibinfo{pages}{014101}
  (\bibinfo{year}{2007}).

\bibitem[{\citenamefont{Essmann et~al.}(1995)\citenamefont{Essmann, Perera,
  Berkowitz, Darden, Lee, and Pedersen}}]{essmann1995smooth}
\bibinfo{author}{\bibfnamefont{U.}~\bibnamefont{Essmann}},
  \bibinfo{author}{\bibfnamefont{L.}~\bibnamefont{Perera}},
  \bibinfo{author}{\bibfnamefont{M.~L.} \bibnamefont{Berkowitz}},
  \bibinfo{author}{\bibfnamefont{T.}~\bibnamefont{Darden}},
  \bibinfo{author}{\bibfnamefont{H.}~\bibnamefont{Lee}}, \bibnamefont{and}
  \bibinfo{author}{\bibfnamefont{L.~G.} \bibnamefont{Pedersen}},
  \bibinfo{journal}{The Journal of Chemical Physics}
  \textbf{\bibinfo{volume}{103}}, \bibinfo{pages}{8577} (\bibinfo{year}{1995}).

\bibitem[{\citenamefont{Lomba et~al.}(2007)\citenamefont{Lomba, Almarza,
  Martin, and McBride}}]{lomba2007phase}
\bibinfo{author}{\bibfnamefont{E.}~\bibnamefont{Lomba}},
  \bibinfo{author}{\bibfnamefont{N.~G.} \bibnamefont{Almarza}},
  \bibinfo{author}{\bibfnamefont{C.}~\bibnamefont{Martin}}, \bibnamefont{and}
  \bibinfo{author}{\bibfnamefont{C.}~\bibnamefont{McBride}},
  \bibinfo{journal}{The Journal of Chemical Physics}
  \textbf{\bibinfo{volume}{126}}, \bibinfo{eid}{244510} (\bibinfo{year}{2007}).

\bibitem[{\citenamefont{Binder}(1982)}]{binder1982scaling}
\bibinfo{author}{\bibfnamefont{K.}~\bibnamefont{Binder}},
  \bibinfo{journal}{Phys. Rev. A} \textbf{\bibinfo{volume}{25}},
  \bibinfo{pages}{1699} (\bibinfo{year}{1982}).

\bibitem[{\citenamefont{Tolman}(1949)}]{tolman1949effect}
\bibinfo{author}{\bibfnamefont{R.~C.} \bibnamefont{Tolman}},
  \bibinfo{journal}{The Journal of Chemical Physics}
  \textbf{\bibinfo{volume}{17}}, \bibinfo{pages}{333} (\bibinfo{year}{1949}).

\bibitem[{\citenamefont{Reichl}(1998)}]{reichl2009modern}
\bibinfo{author}{\bibfnamefont{L.}~\bibnamefont{Reichl}},
  \emph{\bibinfo{title}{A Modern Course in Statistical Physics}}
  (\bibinfo{publisher}{Wiley-Interscience}, \bibinfo{year}{1998}).

\bibitem[{\citenamefont{Muller}(1990)}]{muller1990search}
\bibinfo{author}{\bibfnamefont{N.}~\bibnamefont{Muller}},
  \bibinfo{journal}{Accounts of Chemical Research}
  \textbf{\bibinfo{volume}{23}}, \bibinfo{pages}{23} (\bibinfo{year}{1990}).

\bibitem[{\citenamefont{Graziano and Lee}(2005)}]{graziano2005intactness}
\bibinfo{author}{\bibfnamefont{G.}~\bibnamefont{Graziano}} \bibnamefont{and}
  \bibinfo{author}{\bibfnamefont{B.}~\bibnamefont{Lee}}, \bibinfo{journal}{The
  Journal of Physical Chemistry B} \textbf{\bibinfo{volume}{109}},
  \bibinfo{pages}{8103} (\bibinfo{year}{2005}).

\bibitem[{\citenamefont{Matysiak et~al.}(2011)\citenamefont{Matysiak,
  Debenedetti, and Rossky}}]{matysiak2011dissecting}
\bibinfo{author}{\bibfnamefont{S.}~\bibnamefont{Matysiak}},
  \bibinfo{author}{\bibfnamefont{P.~G.} \bibnamefont{Debenedetti}},
  \bibnamefont{and} \bibinfo{author}{\bibfnamefont{P.~J.}
  \bibnamefont{Rossky}}, \bibinfo{journal}{The Journal of Physical Chemistry B}
  \textbf{\bibinfo{volume}{115}}, \bibinfo{pages}{14859}
  (\bibinfo{year}{2011}).

\end{thebibliography}


\end{document}


\title{Supplementary Material \\ \emph{Temperature and Length Scale Dependence of
    Solvophobic Solvation in a Single-site Water-like Liquid}}

\author{John R. Dowdle, Sergey V. Buldyrev, H. Eugene Stanley, \\ Pablo
  G. Debenedetti, and Peter J. Rossky}

\maketitle

In this supplementary material section we provide details for the Monte Carlo
and molecular dynamics simulations along the saturation curves of the Jagla
fluid, the Lennard-Jones fluid, and the extended simple point charge (SPC/E)
water model.

\begin{table}[!ht]
  \centering
  \begin{tabular}{cc}
    \toprule
    $T$ [$\varepsilon_2 / k_B$] & $P$ [$\varepsilon_2 / r_0^3$] \\
    \midrule
    0.4  &   3.2$\times 10^{-5}$ \\
    0.5  &   5.4$\times 10^{-5}$ \\
    0.6  &   1.4$\times 10^{-4}$ \\
    0.7  &   5.0$\times 10^{-4}$ \\
    0.8  &   1.4$\times 10^{-3}$ \\
    0.9  &   3.0$\times 10^{-3}$ \\
    1.0  &   5.6$\times 10^{-3}$ \\
    1.1  &   9.5$\times 10^{-3}$ \\
    1.2  &   1.5$\times 10^{-2}$ \\
    \bottomrule
  \end{tabular}
  \caption{Pure Jagla liquid $NPT$ MC simulations were performed for several
states along the liquid-vapor coexistence curve. Simulations were performed for
both liquid and vapor densities estimated from the data in Table~1. Each
simulation consisted of 1000 Jagla particles which were simulated for $6\times
10^5$ cycles after being equilibrated for at least $2\times 10^5$ cycles. Each
cycle consists of $N$ MC moves. In each MC move, there is a $1/N$ chance of
attempting a volume move and $(N-1)/N$ chance of attempting to move a randomly
selected particle. Coordinates were output to trajectories every 5 cycles. Test
particle insertion was performed for cavities ranging from 0.5 to
2.0$\sigma_{JG}$ in diameter on each configuration in the liquid trajectories
to obtain the small solute data used in the revised scaled particle theory fit
of Eq. (5). Test particle insertions were performed on vapor trajectories for
all cavity radii listed in Table~\ref{tab:cavity_sims} to obtain non-ideal gas
solubilities and vapor-wall surface tensions \cite{ashbaugh2009blowing}.}
    \label{tab:pure_jagla_sims}
\end{table}

\begin{table}[!ht]
  \centering
  \begin{tabular}{ccc}
    \toprule
    $R$ [$r_0$] & $N$ & $N_{cyc}$ \\
    \midrule
    0.5  &  1000  &  1.76$\times 10^6$  \\
    0.7  &  1000  &  1.76$\times 10^6$  \\
    0.9  &  1000  &  1.76$\times 10^6$  \\
    1.1  &  1000  &  1.76$\times 10^6$  \\
    1.3  &  1000  &  1.76$\times 10^6$  \\
    1.5  &  1000  &  1.76$\times 10^6$  \\
    1.7  &  1000  &  1.76$\times 10^6$  \\
    1.9  &  1000  &  1.76$\times 10^6$  \\
    2.1  &  1000  &  1.76$\times 10^6$  \\
    2.3  &  1000  &  1.76$\times 10^6$  \\
    2.5  &  1000  &  1.76$\times 10^6$  \\
    2.7  &  1000  &  1.76$\times 10^6$  \\
    2.9  &  1000  &  1.76$\times 10^6$  \\
    3.1  &  1000  &  1.76$\times 10^6$  \\
    3.3  &  1000  &  1.76$\times 10^6$  \\
    3.5  &  1000  &  1.76$\times 10^6$  \\
    3.7  &  1000  &  1.76$\times 10^6$  \\
    3.9  &  1000  &  1.76$\times 10^6$  \\
    4.1  &  2000$^*$  &  9.6$\times 10^5$   \\
    4.3  &  2000  &  9.6$\times 10^5$   \\
    4.5  &  2000$^*$  &  9.6$\times 10^5$   \\
    4.7  &  2000  &  9.6$\times 10^5$   \\
    4.9  &  2000  &  9.6$\times 10^5$   \\
    \bottomrule
  \end{tabular}
  \caption{List of $NPT$ MC simulations carried out with $N$ Jagla particles
and a single cavity of radius $R$. For each cavity size, nine simulations were
performed---one for each of the thermodynamic states listed in
Table~\ref{tab:pure_jagla_sims}. The systems were first equilibrated for
$2\times 10^5$ cycles, and Jagla-cavity contact densities were averaged over
$N_{cyc}$ cycles. A superscript $*$ indicates that additional simulations of
$N=3000$ and $N=4000$ Jagla particles were performed to test dependence of the
results on system size. No significant changes were observed.}
  \label{tab:cavity_sims}
\end{table}

\begin{table}[!ht]
  \centering
  \begin{tabular}{cccc}
    \toprule
    $i$ & A$_i$ & B$_i$ & C$_i$ \\
    \midrule
    0 & -0.3233 & 0.6027  & 0.2090 \\
    1 & 1.9374  & -1.2166 & 1.2624 \\
    2 & -1.7246 & 0.0657  & -1.7214 \\
    3 & 0.2920  & 0.4900  & 0.3738 \\
    \bottomrule
  \end{tabular}
  \caption{Cavity equation of state parameters for the Jagla fluid. Parameters
    were obtained from a least squares fit of Eq. (8) to the excess chemical
    potential solvation data in Fig.  5.  Units use $r_0$ for the length scale
    and $\varepsilon_2$ for the energy scale.}
  \label{tab:ceos_params}
\end{table}

\begin{table}[!ht]
  \centering
  \begin{tabular}{cc}
    \toprule
    $T$ [$\varepsilon_{LJ} / k_B$] & $P$ [$\varepsilon_{LJ} / \sigma_{LJ}^3$] \\
    \midrule
    0.65   &   0.0034 \\  
    0.70   &   0.0068 \\  
    0.75   &   0.0096 \\ 
    0.80   &   0.0150 \\ 
    0.85   &   0.0226 \\ 
    0.90   &   0.0327 \\ 
    0.95   &   0.0458 \\ 
    1.00   &   0.0620 \\ 
    \bottomrule
  \end{tabular}
  \caption{Selected states along the liquid-vapor coexistence curve of the LJ liquid \cite{ashbaugh2009blowing}.}
  \label{tab:pure_lj_sims}
\end{table}

\begin{table}[!ht]
  \centering
  \begin{tabular}{cc}
    \toprule
    $T$ [K] & $\rho$ [g/cm$^3$] \\
    \midrule
    273.0  &  1.000 \\ 
    300.0  &  0.998 \\ 
    373.0  &  0.949 \\ 
    423.0  &  0.902 \\ 
    473.0  &  0.841 \\ 
    570.0  &  0.674 \\ 
    610.0  &  0.545 \\ 
    620.0  &  0.507 \\ 
    630.0  &  0.470 \\ 
    640.0  &  0.426 \\ 
    652.0  &  0.326 \\ 
    640.0  &  0.217 \\ 
    630.0  &  0.162 \\ 
    620.0  &  0.112 \\ 
    610.0  &  0.082 \\ 
    570.0  &  0.036 \\ 
    \bottomrule
  \end{tabular}
  \caption{Selected states along the liquid-vapor coexistence curve of SPC/E water \cite{guissani1993coexistence}.}
  \label{tab:pure_spce_sims}
\end{table}

\begin{table}[!ht]
  \centering
  \begin{tabular}{cccccccccc}
    \toprule
    $T$ [K] &  273 & 304 & 335 & 366 & 398 & 429 & 460 & 491 &  522 \\
    \bottomrule
  \end{tabular}
  \caption{Temperatures along the saturation curve of water used in the cavity 
    equation of state calculations.  These are the same reduced temperatures 
    ($T/T_{crit}$, where $T_{crit}$ is the liquid-vapor critical point) as the 
    temperatures used for the Jagla liquid.}
  \label{tab:water_ceos_states}
\end{table}

\clearpage

\begin{table}[!ht]
  \centering
  \begin{tabular}{cccc}
    \toprule
    $i$ & A$_i$ & B$_i$ & C$_i$ \\
    \midrule
    0 &  12.429    & 40.3713   & 12.3712  \\
    1 &  51.3577   & -91.3713  & 19.0438  \\
    2 &  -18.7888  & 28.2881   & -7.68791 \\
    3 &  1.74344   & -2.46828  & 0.735148 \\
    \bottomrule
  \end{tabular}
  \caption{Cavity equation of state parameters for water \cite{ben2005global}. 
    Units use nm for the length scale and kJ/mol for the energy scale.}
  \label{tab:water_ceos_params}
\end{table}

\clearpage

\bibliographystyle{apsrev}
\bibliography{TPL_dep_refs}